\documentclass[10pt,letterpaper]{article}
\usepackage{opex3}
\usepackage{amssymb}
\usepackage{amsmath}
\usepackage{graphicx}
\usepackage{cite}

\begin{document}
\title{Corrected knife-edge-based reconstruction of tightly focused higher order beams}
\author{S. Orlov,$^{*1}$ C. Huber,$^{2,3}$ P. Marchenko,$^{2,3}$ P. Banzer,$^{2,3}$ and G. Leuchs$^{2,3}$} 

\address{
$^1$Center for Physical Sciences and Technology, \\ Savanoriu Ave. 231, LT-02300 Vilnius, Lithuania \\ 
$^2$Max Planck Institute for the Science of Light, \\ G\"{u}nther-Scharowsky-Str. 1, D-91058 Erlangen, Germany \\
$^3$ Institute of Optics, Information and Photonics, University  Erlangen-Nuremberg, \\ Staudtstr. 7B/2, D-91058 Erlangen, Germany \\
}
\email{$^*$sergejus.orlovas@ftmc.lt}

\begin{abstract}
The knife-edge method is an established technique for profiling of even tightly focused light beams. However the straightforward implementation of this method fails if the materials and geometry of the knife-edges are not chosen carefully or in particular if knife-edges are used that are made of pure materials. In these cases artifacts are introduced in the shape and position of the reconstructed beam profile due to the interaction of the light beam under study with the knife. Hence, corrections to the standard knife-edge evaluation method are required. Here we investigate the knife-edge method for highly focused radially and azimuthally polarized beams and their linearly polarized constituents. We introduce relative shifts for those constituents and report on the consistency with the case of a linearly polarized Gaussian beam. An adapted knife-edge reconstruction technique is presented and proof-of-concept tests demonstrating the reconstruction of beam profiles are shown.
\end{abstract}

\section{Introduction}
The circular shape of a paraxial linearly polarized Gaussian beam undergoes, upon high numerical aperture focusing, an elongation of its focal spot along the polarization axis of the input beam \cite{SQuab00,RDorn03a,BRic59}. Upon tight focusing radially and azimuthally polarized beams produce not only highly confined but also symmetric electric and magnetic field distributions in the focal plane and a strong longitudinal electric field component on the optical axis for radial polarization \cite{RDorn03a,KSYoug00,RDorn03b}. A radially polarized mode can be decomposed into two orthogonally polarized Hermite-Gaussian (HG) modes: an $x$-polarized HG$_{10}$ and a $y$-polarized HG$_{01}$ mode. In contrast, an azimuthally polarized mode is a superposition of a $y$-polarized HG$_{10}$ and an $x$-polarized HG$_{01}$ mode. Similar to linearly polarized Gaussian beams, the tight focusing of these linearly polarized constituents also results in a symmetry break of the focal spot. Due to the rich structures of the focal spot achieved by various engineering techniques a precise measurement of such complicated vectorial fields is rather challenging. Nevertheless it is crucial to experimentally analyze and profile tightly focused vectorial beams in a real-world setup before utilizing them for experiments in nano-optics (see \cite{JKin07,Zuech08,PBan10,PBan102,TBau2015} and others).

In literature many methods for beam characterization are described such as the knife-edge \cite{AHFire77,JMKho83,MACdeA09,GBro85}, point scan method  \cite{MBSch81} or slit method \cite{RLMcC84}. In the knife-edge method, a beam-block realized by a sharp edge made from an opaque material (such as a knife- or razor-blade) is line-scanned through the beam perpendicular to its optical axis while the transmitted power is monitored by a detector for several scanning directions. From the resulting photocurrent curves (power vs. position of the edge relative to the beam) the so-called beam-projections onto the scanning-line and finally the beam shape can be tomographically reconstructed \cite{RDorn03a, RDorn03b}.

In a more recent study, knife-edges made from pure materials (metals, etc.) were systematically studied and polarization dependent effects in the knife-edge profiling method were observed. Those effects result in a shift and a deformation of the measured projections and depend on the polarization and wavelength of the input beam and the materials of the knife-edge samples. Caused by the aforementioned distortions introduced by knife-edges made from pure materials, a proper reconstruction of the beam under study seems to be impossible using a standard evaluation method \cite{PMar11}. Recently, we have demonstrated, that the interaction between the knife-edge and a highly focused linearly polarized beam can be understood in terms of the moments of the beam (beam profile times a polynomial) and, therefore, an adapted beam reconstruction and fitting technique can be successfully applied in this case \cite{CHub13}.

The aim of the study discussed below is to apply the knowledge already obtained for linearly polarized beams \cite{PMar11, CHub13}  and to develop an adapted knife-edge reconstruction technique for highly focused radially and azimuthally polarized beams and their linearly polarized constituents.

\section{Theoretical considerations}
\subsection{Basics of the knife-edge method}

\begin{figure}[h!t!]
\centering
\includegraphics[scale=0.35]{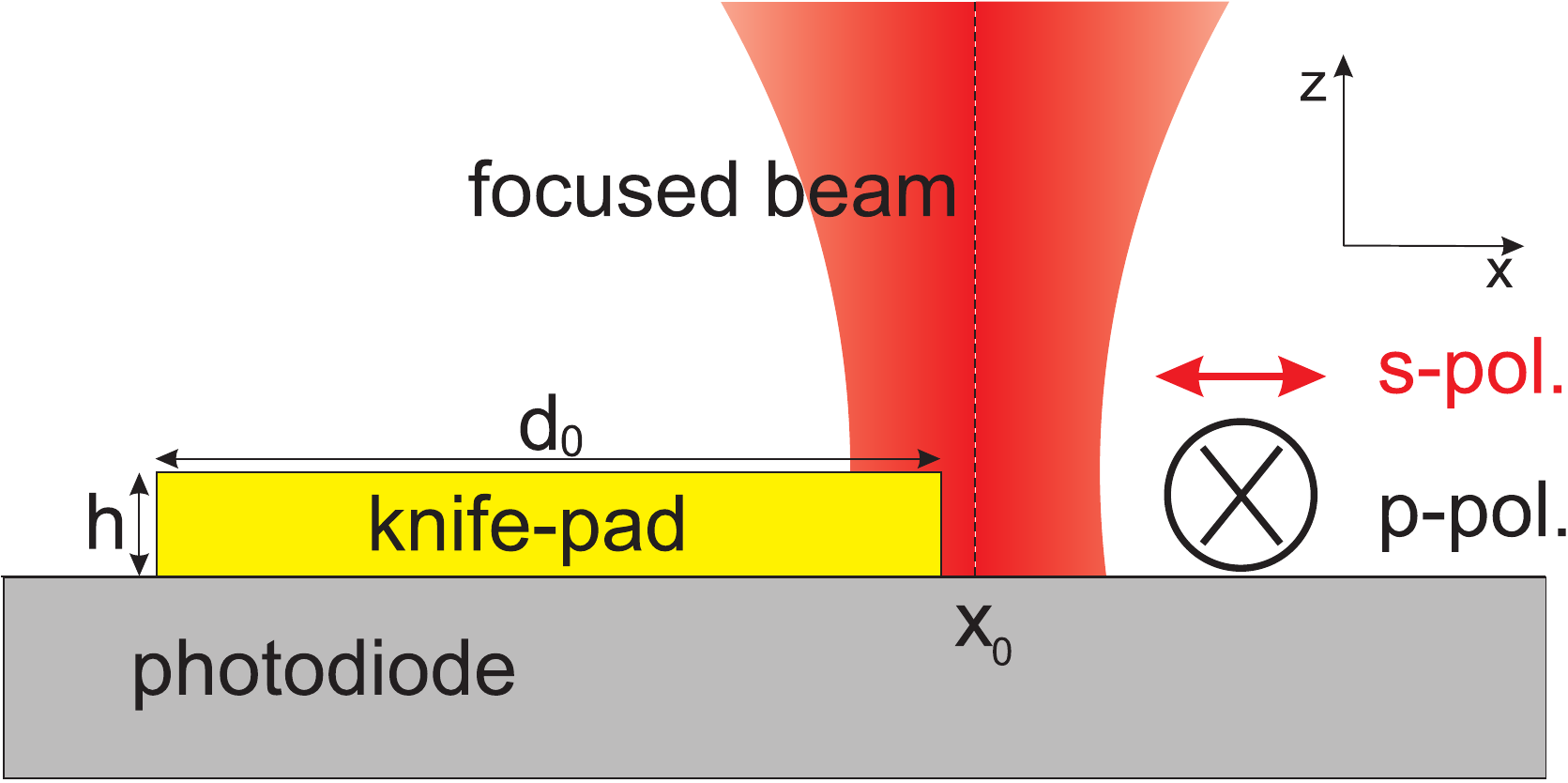}\text{(a)} 
\includegraphics[scale=0.35]{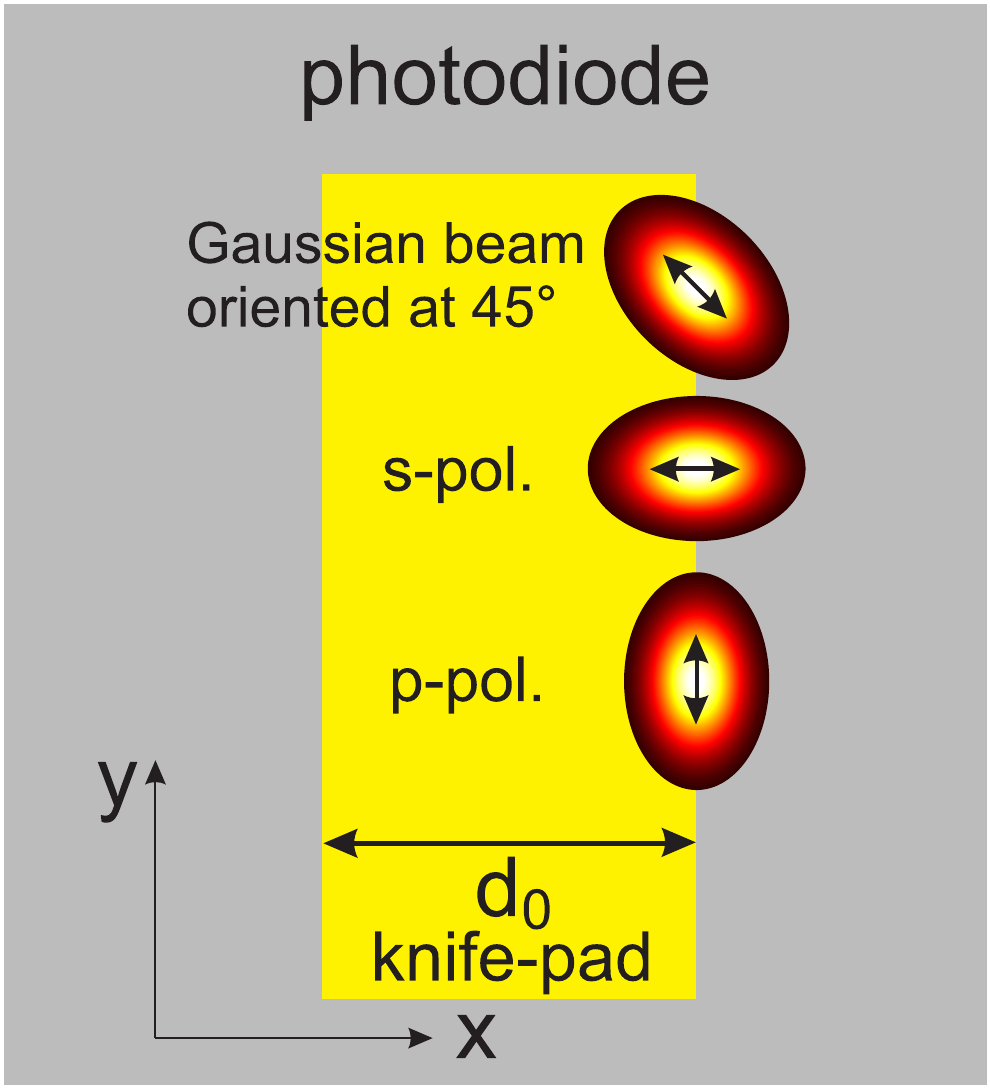}\text{(b)} 
\includegraphics[scale=0.19]{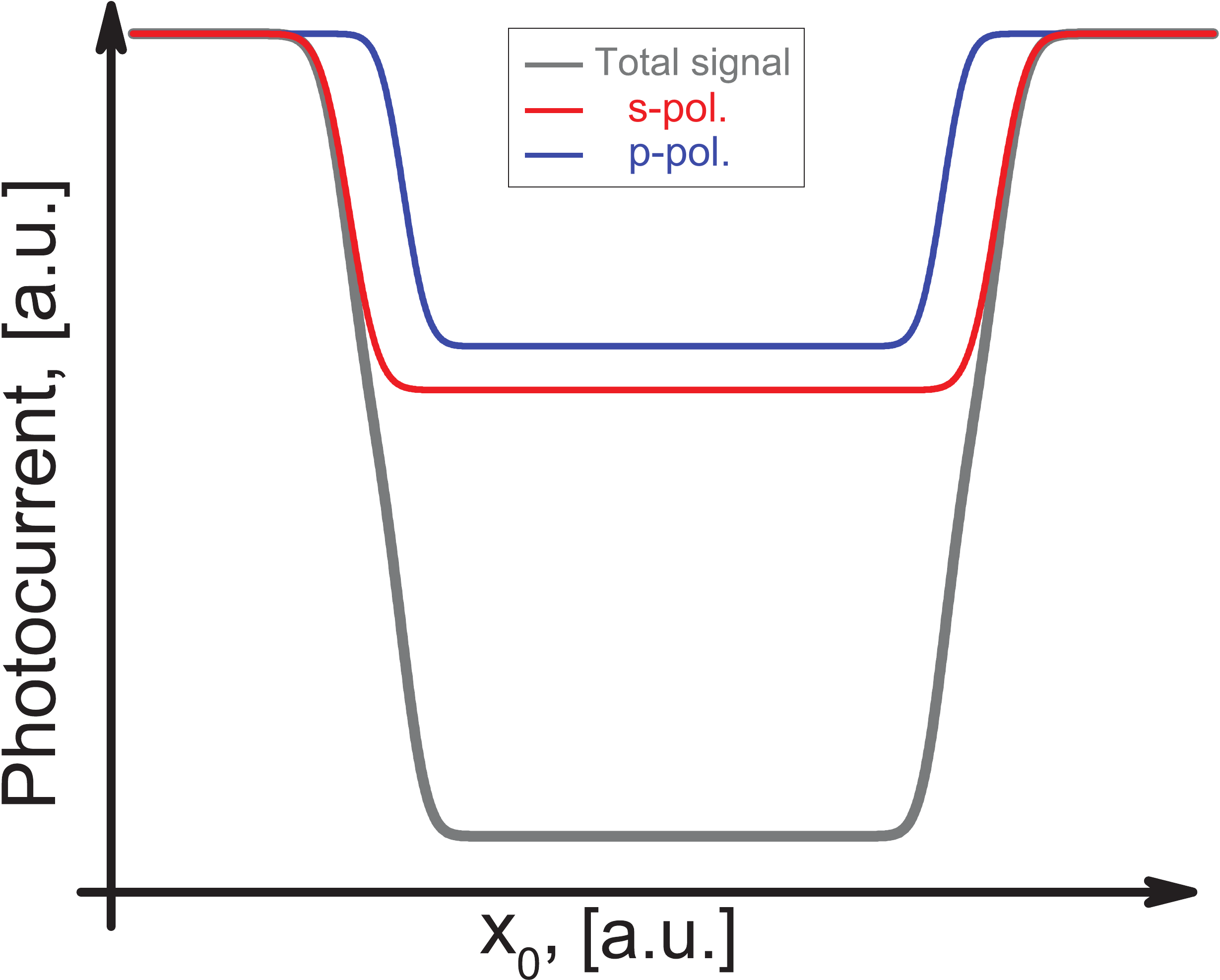}\text{(c)}
\includegraphics[scale=0.19]{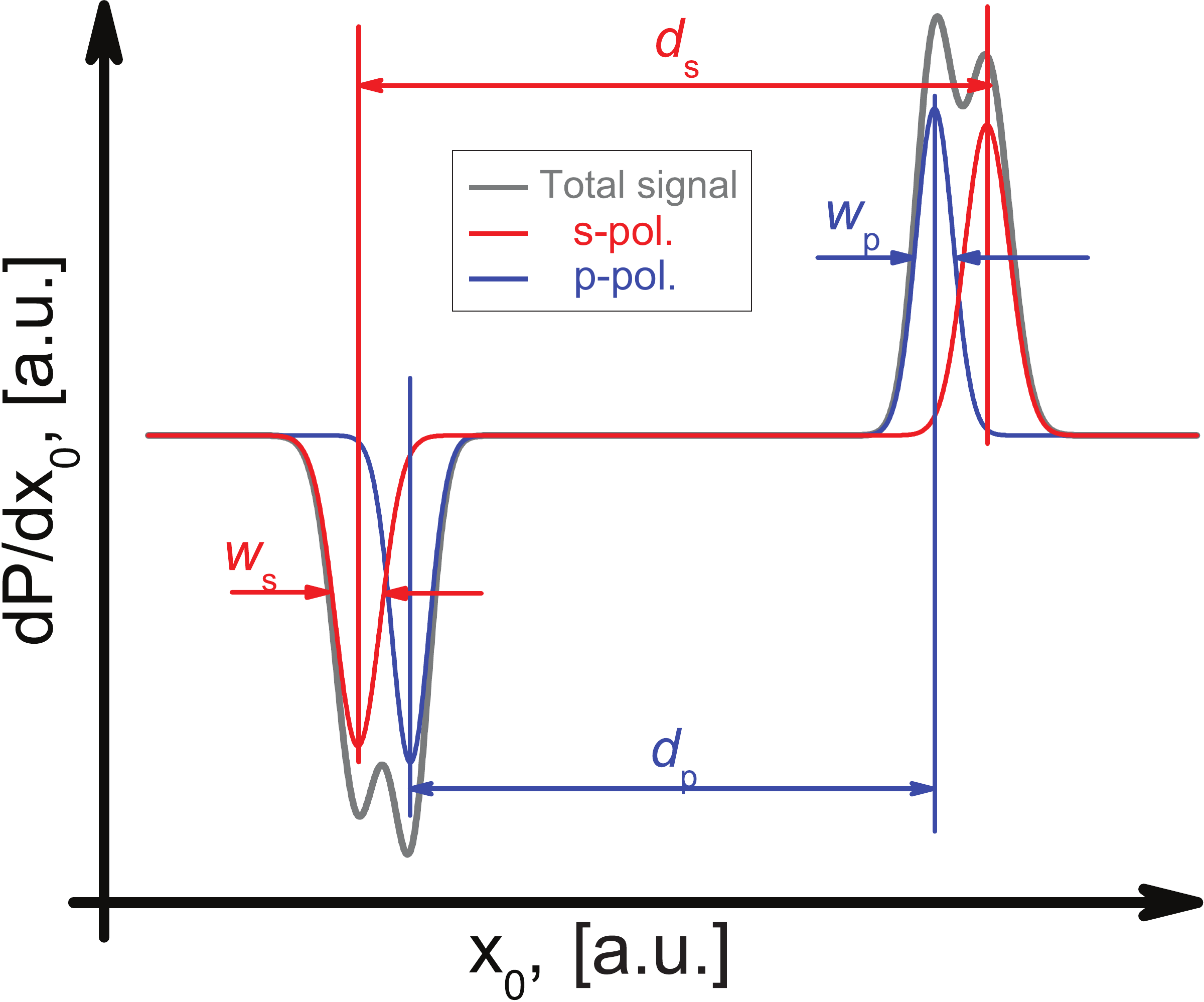}\text{(d)}
\caption{Schematic depiction of the knife-edge method $xz$-plane (a) and $xy$-plane (b). The state of polarization always refers to the orientation of the electric field of the incoming beam relative to the knife-edge in the $xy$-plane. Typical beam profiling data (photocurrent curves) c) and their derivatives (beam-projections) (d) for a linearly polarized Gaussian beam (oriented at $45$ degrees (total signal) - gray, $s$-polarization - red, $p$-polarization - blue).}
\label{fig:exp_setup}
\end{figure}

The principle of the knife-edge method is depicted in Fig. \ref{fig:exp_setup}. For experimental reasons we consider here beam profiling by two adjacent edges of a single rectangular metal knife-pad. The photocurrent generated inside the photodiode is proportional to the power $P$ detected by the photodiode and is recorded for each beam position $x_0$ with respect to the knife-edge
\begin{equation}
 P=P_0\int _{-\infty}^{\infty}\mathrm{d}y \int _{-\infty}^0  I\left(x+x_0,y,z=0 \right) \mathrm{d}x,
\label{eq:knife}
\end{equation}
where $P_0$ is a proportionality coefficient and $I$ is the electric field intensity. In the conventional knife-edge method the derivative $\partial P/\partial x_0$ of the photocurrent curve with respect to the beam position $x_0$ reconstructs a projection of the intensity onto the $xz$-plane at $z=0$ (projection onto the $x$-axis) \cite{AHFire77}. In a next step, the two dimensional electric field intensity distribution can be reconstructed from projections measured along different directions using the Radon back-transform, if polarization dependent effects can be neglected \cite{RDorn03a, RDorn03b}.

In this context it has to be mentioned that the term intensity usually refers to the total electric energy density and at the same time to the $z$-component of the Poynting-vector $\mathbf{S}$, because they are proportional to each other due to longitudinal electric field components being negligible in the limit of paraxial light beams. In the case of tightly focused light beams (non-paraxial propagation), electric fields can exhibit strong longitudinal electric field components resulting in different distributions of $|\mathbf{E}(x,y)|^2$ and $S_z (x,y)$. It was shown that the integral equation (\ref{eq:knife}) adopted from the conventional knife-edge method allows for the reconstruction of the beam profile in terms of its total electric energy density distribution $|\mathbf{E}(x,y)|^2$ also in case of tightly focused vectorial beams if special edge-materials, thicknesses and certain wavelengths are chosen \cite{RDorn03a}. Nevertheless, with pure knife-edge materials of different thicknesses and for different wavelengths of the input beam, the retrieved projections do not correspond to the expected projections of the electric energy density distributions as they appear strongly distorted \cite{PMar11}.

To better understand those shifts and distortions in the measured projections we exemplarily use a fundamental Gaussian beam, which has its electric field oriented at $45$ degrees to the knife-edge. With a polarizer in front of the focusing objective oriented either perpendicular or parallel to the knife-edge, the incident beam is then decomposed in its linearly polarized constituents with the electric field of the incoming beam being perpendicular ($s$) or parallel ($p$) to the knife-edge, see Figs. \ref{fig:exp_setup}(a) and \ref{fig:exp_setup}(b). A photocurrent curve is recorded for each orientation of the polarizer and the sum of those two photocurrent curves should result in the signal recorded without the polarizer (total signal), see Fig. \ref{fig:exp_setup}(c). Taking the derivative $\partial P/\partial x_0$ of the photocurrent curves with respect to the beam position $x_0$ should reconstruct the expected Gaussian beam profile, however this is not the case (see, Fig. \ref{fig:exp_setup}(d)). The projections of the Gaussian beam with its electric field oriented at $45$ degrees to the knife-edge is strongly distorted. They exhibit in a exaggerated picture two lobes with a minima between the lobes. However, when a polarizer is used the situation looks different: each linear constituent preserves in principle its Gaussian shape and only smaller distortions can be found. It becomes evident, that $s$- and $p$-projections are shifted ($d_s \neq d_p \neq d_0$ with $d_0$ the width of the metal pad (see Fig. \ref{fig:exp_setup}(a))) and asymmetrically deformed also causing deviations in the retrieved beam diameters $w_s$  and $w_p$ (see Fig. \ref{fig:exp_setup}(d)). The above-mentioned effects are caused by the fact that the knife-edge is not only blocking the beam while line-scanning but it is also excited by the beam plasmonically. Furthermore, the power flow through the knife-edge is polarization dependent and proportional to the value of the projection of the electric energy density onto the edge \cite{PMar11, CHub16}. Obviously, if one does not account for these effects, the standard scheme is not valid without corrections unless the knife-edge parameters are carefully chosen \cite{SQuab00}. It is worth noting here that during the reconstruction of light beams with diameters larger than several wavelengths such effects are negligible as the distortions are much smaller than the projection of the beam.

\begin{figure}[h!t!]
\centering
\includegraphics[scale=0.19]{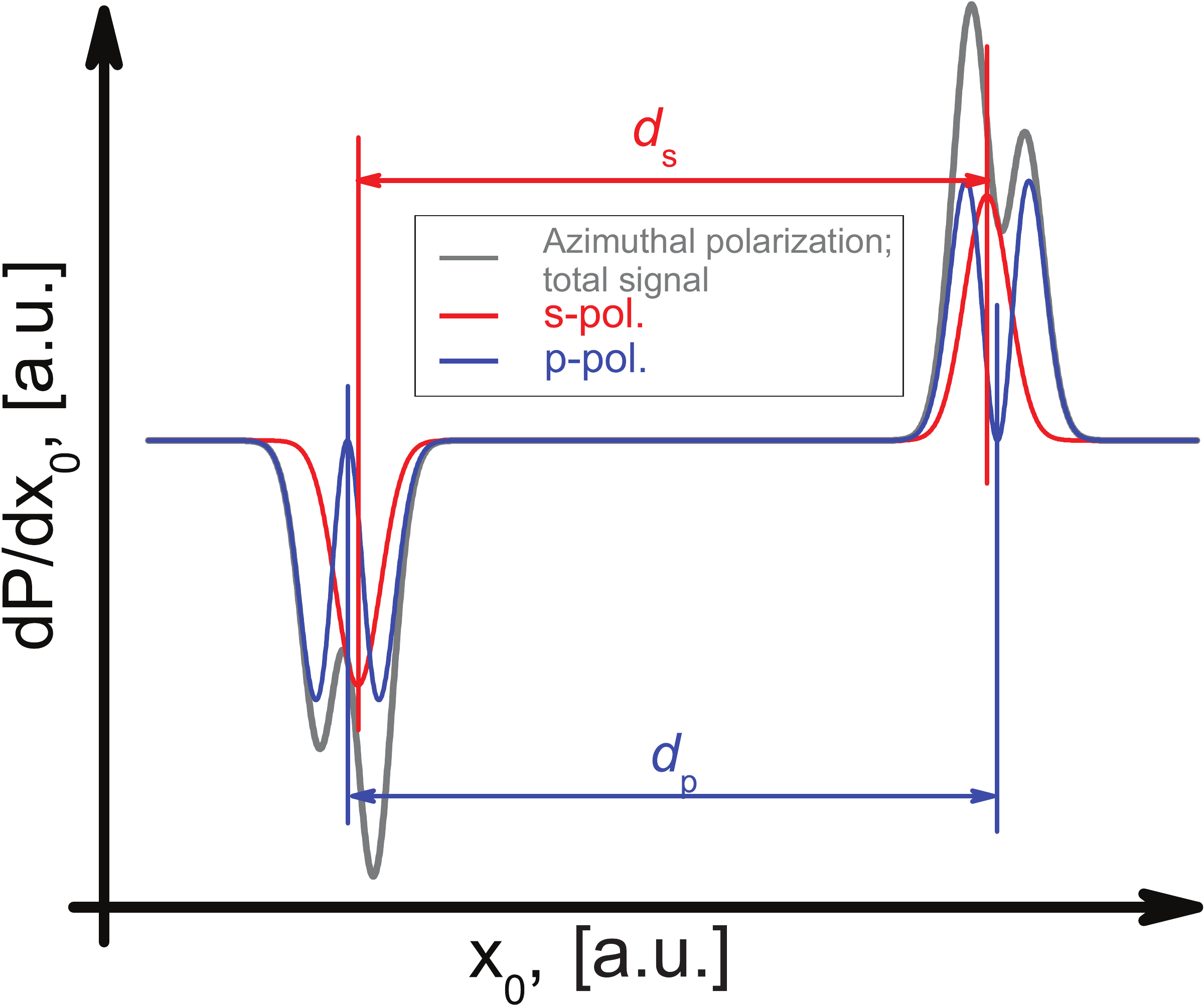}\text{(a)}
\includegraphics[scale=0.19]{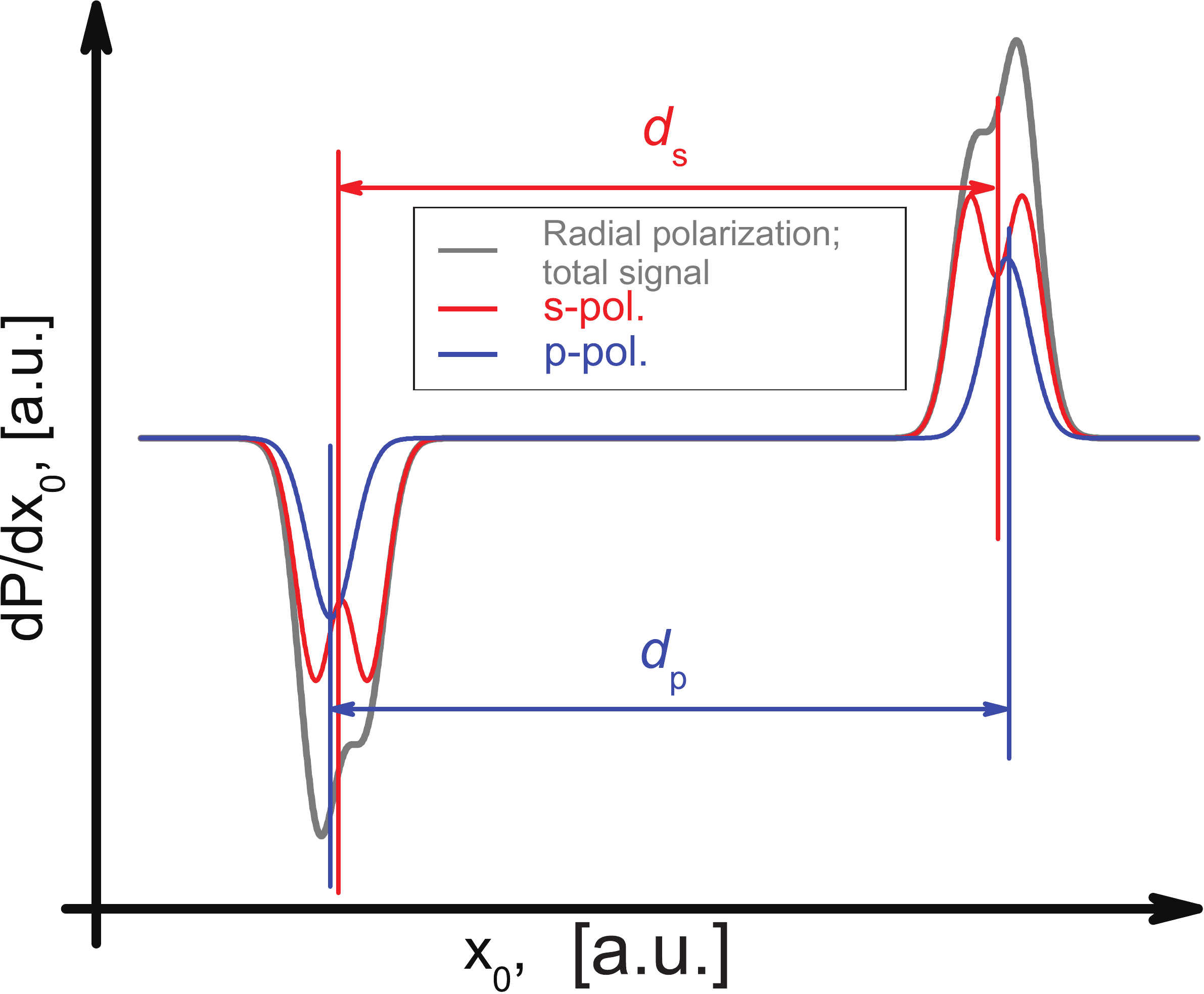}\text{(b)}
\caption{Typical derivatives of beam profiling data (photocurrent curves) for azimuthally (a) and radially (b) polarized beams (gray) and their linear constituents  ($s$-polarization - red, $p$-polarization - blue).}
\label{fig:exp_setup2}
\end{figure}

We would like to particularly discuss now a similar problem, which is observed while beam-profiling tightly focused azimuthally or radially polarized beams, see Fig. \ref{fig:exp_setup2}. Compared to the rotated linearly polarized Gaussian beam a radially polarized beam is also a superposition of two orthogonally polarized modes ($x$-pol. HG$_{10}$, $y$-pol. HG$_{01}$). In a similar fashion an azimuthally polarized beam is also a superposition of two linearly polarized constituents ($y$-pol. HG$_{10}$, $x$-pol. HG$_{01}$). Also in the case of tightly focused cylindrical vector beams the retrieved total beam projections will be strongly modified, if their linearly polarized constituents are shifted and distorted, see Figs. \ref{fig:exp_setup2}(a) and \ref{fig:exp_setup2}(b). Therefore it is important to investigate whether the distortions observed in these linearly polarized higher-order constituents can also be described using analogical parameters $d_s$ and $d_p$, as in the case of a fundamental linearly polarized Gaussian beam \cite{PMar11}. Furthermore, it is crucial to study whether the apparent shifts $d_s$, $d_p$ of the projections correlate with the previously investigated case.

Additionally, we also discuss the applicability of an adapted reconstruction method, discussed and introduced for tightly focused fundamental Gaussian beams recently \cite{CHub13}.

\subsection{Approximation of vector beams via paraxial modes}
We start with a discussion of the basis functions that would be most suitable for the further development of an adapted knife-edge technique. An accurate description of highly focused beams is possible either by vectorial diffraction theory, leading to a numerical calculation of integrals, see for instance \cite{BRic59, KSYoug00}, or by the so-called complex source approach, which enables an analytical description of variously polarized tightly focused fields \cite{SOrl10, SOrl14}. The disadvantage of both approaches is their complexity, which makes them unsuitable for the development of an adapted knife-edge technique. Our aim here is to use an orthogonal set of functions, which would be sufficient for an approximate description of highly focused fields. Electric (and magnetic) field components that are a solutions to Maxwell`s equations can be expressed in terms of two independent functions $f_1\left(\mathbf{r} \right)$ ($x$-polarized) and $f_2\left(\mathbf{r} \right)$ ($y$-polarized) of the paraxial wave equation \cite{WLEri94}:
\begin{align}
E_x = & f_1\left(\mathbf{r} \right) +\frac{1}{4k^2}\left(\frac{\partial ^2 f_1\left(\mathbf{r} \right)}{\partial x^2} +\frac{\partial ^2 f_1\left(\mathbf{r} \right)}{\partial y^2} \right) + \frac{1}{2k^2}\frac{\partial ^2 f_2\left(\mathbf{r} \right)}{\partial x \partial y}, \nonumber \\
E_y = & f_2\left(\mathbf{r} \right) -\frac{1}{4k^2}\left(\frac{\partial ^2 f_2\left(\mathbf{r} \right)}{\partial x^2} +\frac{\partial ^2 f_2\left(\mathbf{r} \right)}{\partial y^2} \right) + \frac{1}{2k^2}\frac{\partial ^2 f_1\left(\mathbf{r} \right)}{\partial x \partial y}, \nonumber \\
E_z = & \frac{\mathrm{i}}{k} \left(\frac{\partial  f_1\left(\mathbf{r} \right)}{\partial x} +\frac{\partial  f_2\left(\mathbf{r} \right)}{\partial y} \right),
	\label{eq:MaxGauss}
	\end{align}
	
We can further simplify these expressions by keeping only the leading terms in $f_1\left(\mathbf{r} \right)$ and $f_2\left(\mathbf{r} \right)$, i.e. we drop second derivatives from expressions (\ref{eq:MaxGauss}), if a lower order derivative appears there. This can be done because the second derivatives are of order $1/k^2l^2_0$, where $l_0$ is some characteristic length. Next, we identify functions $f_1\left(\mathbf{r} \right)$ and $f_2\left(\mathbf{r} \right)$ as the $x$- and $y$-polarized constituents of the incident beam. Furthermore we introduce the so-called elegant HG modes \cite{AESie86}:

\begin{align}
f^{(m,n)}_{1,2}\left(\mathbf{r} \right)=\sigma^{\left(m+n \right)/2+1} H_m\left(x\sigma\right) H_n\left (y\sigma\right) \exp \left[\mathrm{i}kz -\sigma^2\left(x^2+y^2\right) -\mathrm{i}\left(1 + \frac{m+n}{2} \right)\arctan \xi \right],
	\label{eq:eHG}
	\end{align}
with
\begin{align}
\sigma = \frac{1}{\omega_0\sqrt{1+ \mathrm{i}\xi }}, \quad \xi = z/z_0, \quad z_0=k\omega^2_0/2,
	\label{eq:sig_xi}
	\end{align}
and $\omega_0$ the beam width. We note here, that though single Hermite-Gaussian modes are not exact solutions to Maxwell`s equations, corrections can be found and expressed as an infinite sum according to the method of Lax et al. \cite{Lax75} and can be related to complex sourced vortices \cite{SOrl14}. The main advantage of using the elegant version of Hermite-Gaussian modes over standard modes is the following useful relation \cite{AESie86}
\begin{align}
f^{(m,n)}_{1,2}\left(\mathbf{r} \right) = \frac{\partial^{m+n}f^{(0,0)}_{1,2}\left(\mathbf{r} \right)}{\partial x^m \partial y^n},
	\label{eq:fmn3}
	\end{align}
which greatly simplifies further considerations and enables us to rewrite Eq. (\ref{eq:MaxGauss}) as
 \begin{align}
E_x\left(\mathbf{r} \right) = \sum _{m=m_1}^{m_2}\sum _{n=n_1}^{n_2}a_{m,n}f^{(m,n)}_{1}\left(\mathbf{r} \right) + \frac{1}{2k^{2}}\sum _{m=m_3}^{m_4}\sum _{n=n_3}^{n_4}b_{m,n}f^{(m+1,n+1)}_{2}\left(\mathbf{r} \right), \nonumber \\
E_y\left(\mathbf{r} \right) = \frac{1}{2k^2}\sum _{m=m_1}^{m_2}\sum _{n=n_1}^{n_2}a_{m,n}f^{(m+1,n+1)}_{1}\left(\mathbf{r} \right) + \sum _{m=m_3}^{m_4}\sum _{n=n_3}^{n_4}b_{m,n}f^{(m,n)}_{2}\left(\mathbf{r} \right), \nonumber \\
E_z\left(\mathbf{r} \right) = \frac{\mathrm{i}}{k}\sum _{m=m_1}^{m_2}\sum _{n=n_1}^{n_2}a_{m,n}f^{(m+1,n)}_{1}\left(\mathbf{r} \right) +  \frac{\mathrm{i}}{k}\sum _{m=m_3}^{m_4}\sum _{n=n_3}^{n_4}b_{m,n}f^{(m,n+1)}_{2}\left(\mathbf{r} \right) ,
	\label{eq:sumEyz}
	\end{align} 
	
where expansion coefficients $a_{m,n}$ and $b_{m,n}$ uniquely describe a highly focused field and its linearly polarized constituents. In a similar fashion we can express the projections of electric field densities for focal planes of $s$- or $p$-polarized projections of arbitrary beams. We multiply Eq. (\ref{eq:sumEyz}) with its complex conjugate and integrate over the whole $y$-axis, so indices $n$ disappear from the sum (6). In this manner we arrive at the following expressions for $p$- and $s$-polarized projections
	
	\begin{align}
	U_{E,x}\left(x \right) = \sum _{m_1,m_2}B_{m_1,m_2}g^{(m_1)}_{1}\left( x\right)g^{(m_2)}_{1}\left(x \right) , \nonumber \\
	U_{E,y}\left(x \right) = \sum _{m_1,m_2}B_{m_1,m_2}g^{(m_1)}_{2}\left( x\right)g^{(m_2)}_{2}\left(x \right), \nonumber \\
	U_{E,z}\left(x \right) = \sum _{m_1,m_2} \alpha _{m_1,m_2}B_{m_1,m_2}g^{(m_1+1)}_{1}\left( x\right)g^{(m_2+1)}_{1}\left(x \right),
	\label{eq:sumU}
	\end{align}

\begin{figure}[t!]
\centering
\includegraphics[scale=0.16]{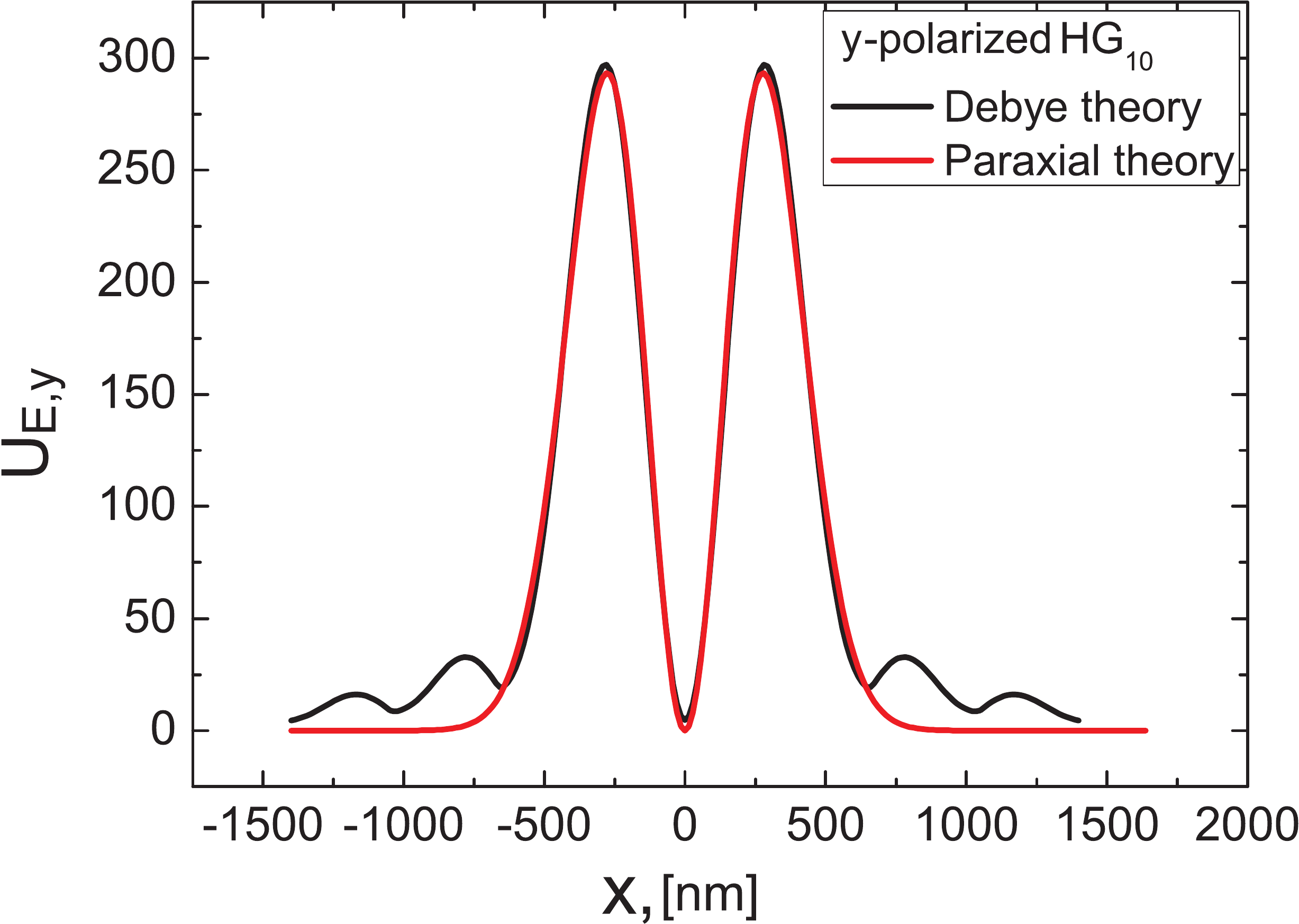}\text{(a)}
\includegraphics[scale=0.16]{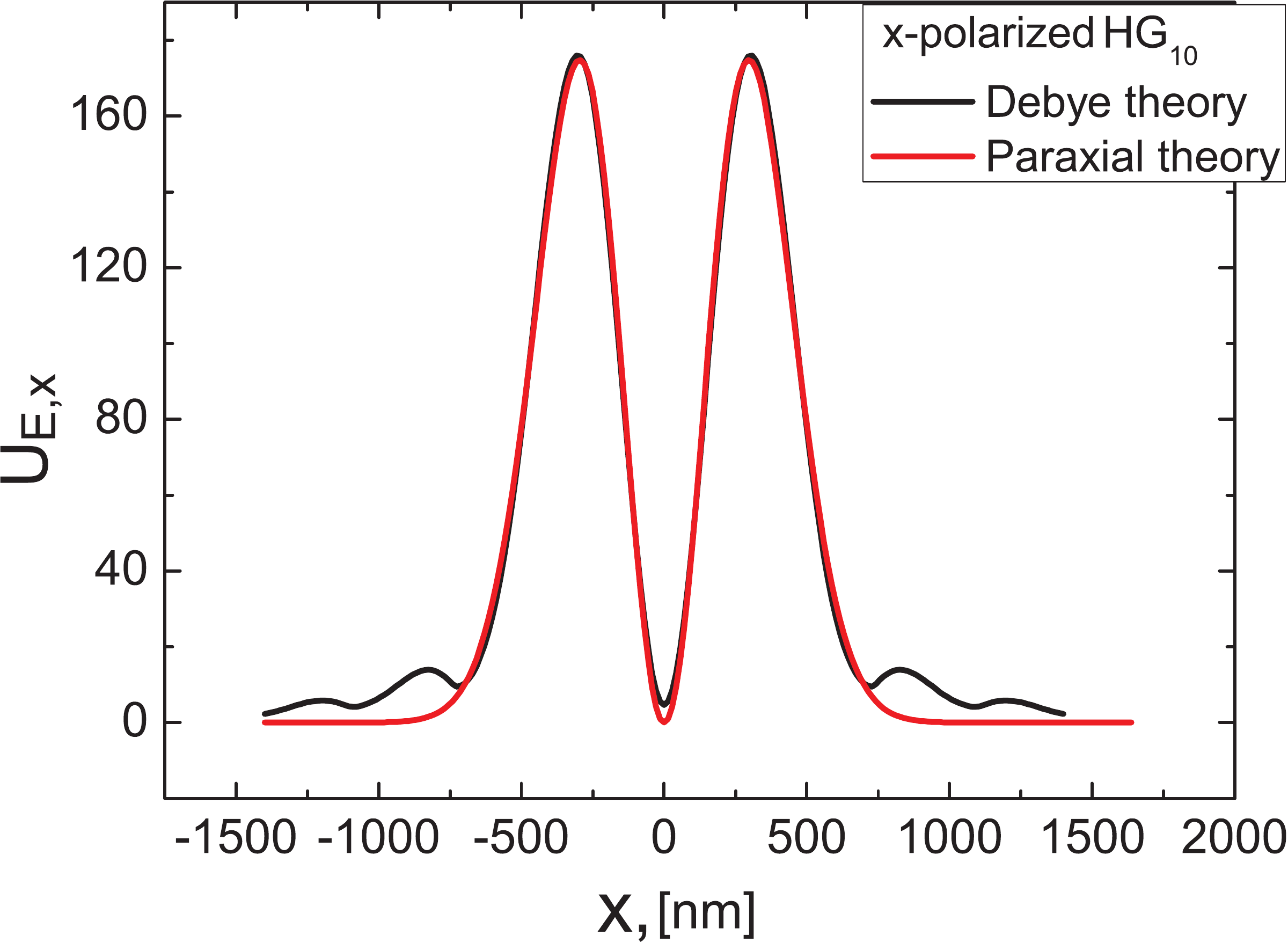}\text{(b)}
\includegraphics[scale=0.16]{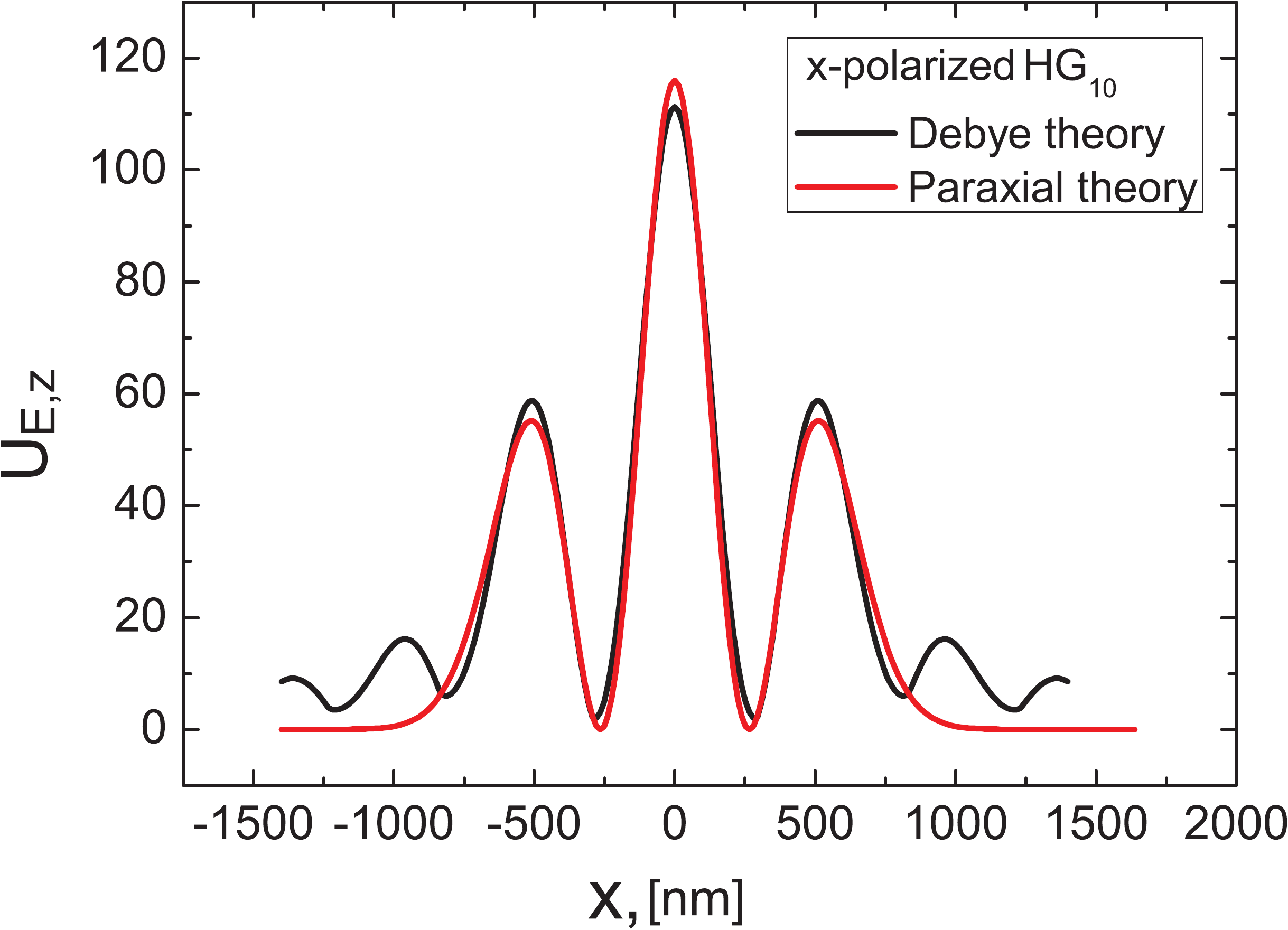}\text{(c)}
\includegraphics[scale=0.16]{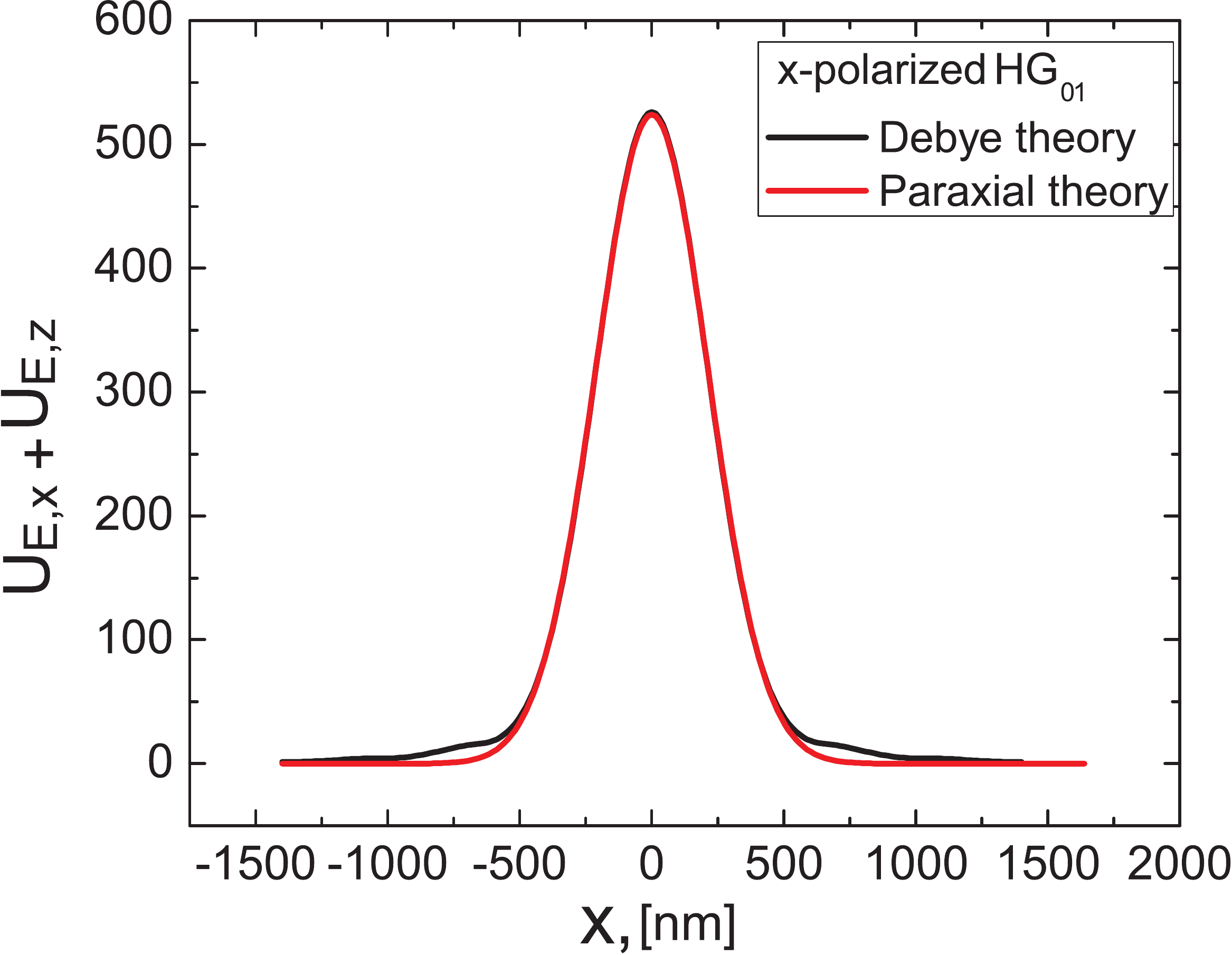}\text{(d)}
\includegraphics[scale=0.16]{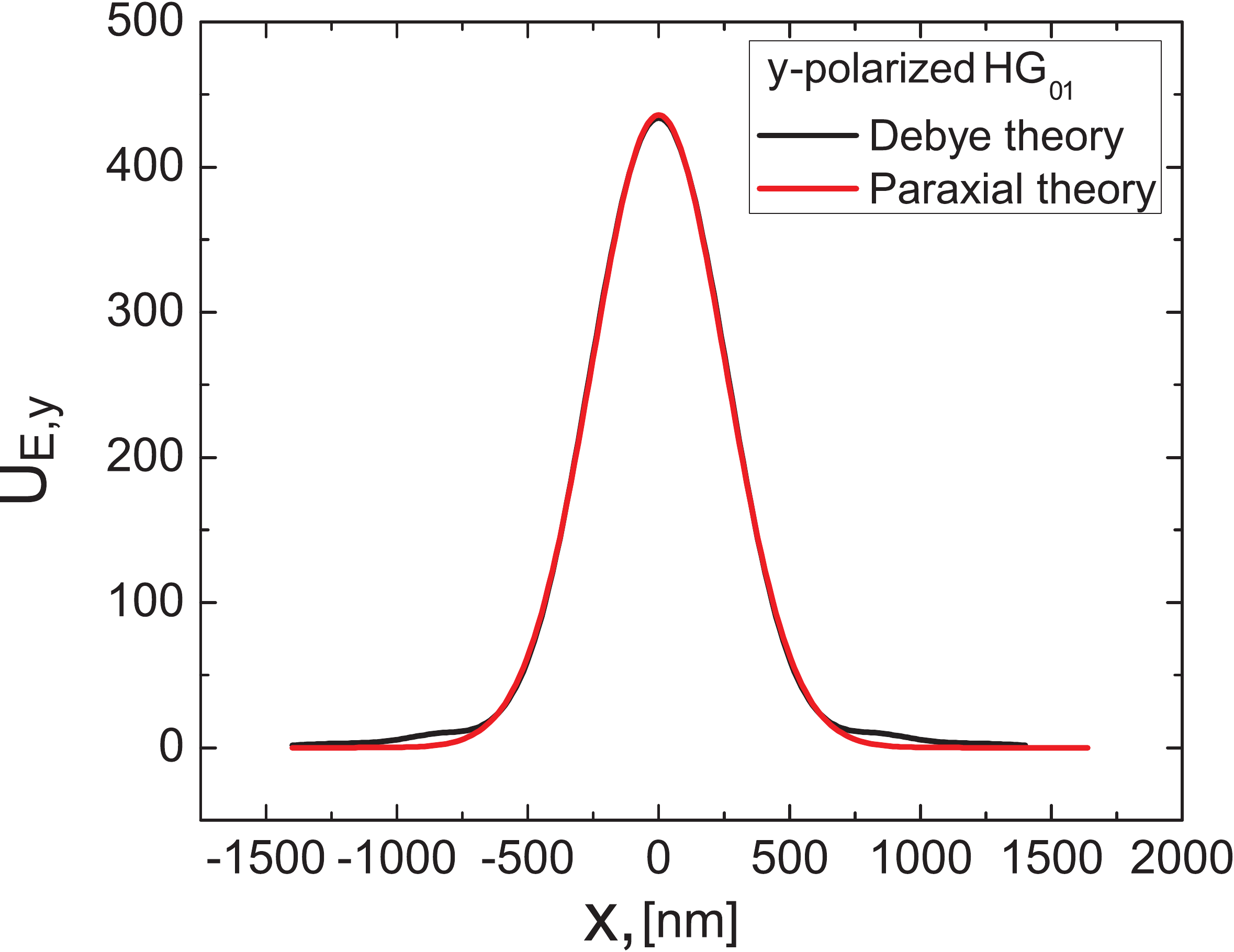}\text{(e)}
\caption{Comparison of calculated components of the projections of electric field densities between an approximation (paraxial theory, red) and exactly calculated projections using Richards-Wolf integrals (vectorial diffraction theory (Debye), black). The wavelength is $\lambda = 700$ nm, the numerical aperture ($NA$) is $0.9$, the focal length is $f=2.0$ mm and the beam width of HG modes at the entrance pupil is $w_0=1.74$ mm. Indices for Hermite-Gaussian modes are shown in the graphs.}
\label{fig:HGmodes}
\end{figure}
	
where $B_{m_1,m_2}$ are unknowns describing intensity profiles, functions $g_{1,2}^{(m)}$ are projections of HG-modes $f^{(m,n)}_{1,2}$ onto the $xz$-plane and $\alpha _{m_1,m_2}$ are coefficients, which correct the amplitudes of $z$-components in Eq. (\ref{eq:sumEyz}) and have be obtained separately.

Let us now discuss an azimuthally and radially polarized beam and their linear constituents in more detail. A paraxial radially polarized beam is a superposition of two orthogonal HG modes, an x-polarized $f^{(1,0)}_1$ and a y-polarized $f^{(0,1)}_2$ mode, whereas a paraxial azimuthally polarized beam is obtained by a superposition of a y-polarized  $f^{(1,0)}_2$ and an x-polarized $f^{(0,1)}_1$mode. Under tight focusing, each HG-mode undergoes changes and new components of the electric field appear, which can be described using Eq. (\ref{eq:MaxGauss}). However we restrict ourselves here to the approximations of Eq. (\ref{eq:sumEyz}), which we use to determine the shape of projections in the form of Eq. (\ref{eq:sumU}). We compare now projections, which we derive using Eq. (\ref{eq:sumU}), with projections calculated by Richards-Wolf integrals \cite{BRic59} for cases which we consider in the experimental part. For one particular wavelength we demonstrate the outcome of a fitting procedure for several projections of Hermite-Gaussian input beams in Fig. \ref{fig:HGmodes}, where we used as a criterion for convergence that the paraxial beam overlaps with the central part of the exactly obtained projections. In this way the central part of the $p$-polarized projection of a HG$_{10}$ beam ($y$-polarized HG$_{10}$) was fitted by a paraxial beam function from Eq. (\ref{eq:sumU}) and the fitting outcome is shown in Fig. \ref{fig:HGmodes}(a). As a result we can see that the paraxial model describes the $y$-component sufficiently well. The central part of the $s$-polarized projection of a HG$_{10}$ beam ($x$-polarized HG$_{10}$) was also fitted using paraxial beam functions, see Figs. \ref{fig:HGmodes}(b) and \ref{fig:HGmodes}(c). Here we can seen that the $x$-component can be approximated well, whereas the $z$-component shows minor discrepancies in the height of the intensities. Lastly, we have fitted the $s$- and $p$-polarized projection of a HG$_{01}$ beam ($x$-, $y$-polarized HG$_{01}$) using paraxial beam functions from Eq. (\ref{eq:sumU}) and the results presented in Figs. \ref{fig:HGmodes}(d) and \ref{fig:HGmodes}(e) showing the good quality of the approximation.

\subsection{Corrections to the knife-edge based reconstruction scheme for higher order modes}
In our previous work \cite{CHub13}, we have presented a numerical technique to correct for artifacts in profiling of linearly polarized fundamental Gaussian beams, which are introduced by the interaction of the knife-edge with the focused light field \cite{CHub16}. This approach finally enables the use of any kind of opaque material as knife-edge. Now we discuss a generalization of that numerical technique, which will also allow for the correction of artifacts observed in beam profiling of $s$- and $p$-polarized projections of the electric field, which can be represented using Eq. (\ref{eq:sumU}). For that purpose, we start a short discussion about light-matter interaction between the focused light beam and the knife-edge as it is recorded by a detector, see Eq. (\ref{fig:exp_setup}).

First, as we have already noticed, the integration in Eq. (\ref{eq:knife}) over the $y$-axis reduces the dimensionality of the electric field density. Therefore beam profiling does not result in reconstruction of the beam intensity $I$ but its projection $U_E$ onto the $xz$-plane. Thus, eigenmodes of the knife-edge problem consist of two independent classes: transverse electric (in our notation $p$-polarized) and transverse magnetic ($s$-polarized) modes. The projection of the electric field density $U_{E}$ in the $p$-case has a non-vanishing component of the electric field density $U_{E,y}$ parallel to the knife-edge and a $z$-component $U_{E,z}$, with an indistinguishable shape from $U_{E,y}$ due to the symmetry of Eq. (\ref{eq:sumU}). The $s$-modes have two non-vanishing and distinguishable components of the electric field density, where the main component $U_{E,x}$ is perpendicular ($s$-polarization) to the knife-edge \cite{PMar11, BStur07, SEKoc09}. 
In order to analyze the interaction of the electric field distribution $E_b$ of a highly focused beam with the knife-edge, we need to start by decomposing it into its $s$- and $p$-polarized constituents. In this manner the resulting beam will be described by a sum (Eq. (\ref{eq:sumEyz})). We start with taking the derivative of Eq. (\ref{eq:knife}) and rewriting the result as \cite{CHub13, CHub16}
\begin{align}
	\frac{\partial P}{P_0\partial x_0}=U_{E}(x_0) + \sum_{n=1}^{\infty}C_n\frac{\partial^n U_{E}( x_0)}{\partial x_0^n},
	\label{eq:adopt_der}
\end{align}
with $C_n = (\mathrm{i}^nn!)^{-1}\partial^n \hat{T}/\partial k^n_x$. Here $U_E(x)$ is the projection of the electric field energy density onto the $xz$-plane at the knife-edge and $\hat{T}(k_x)$ is a spectral representation of the polarization dependent knife-edge interaction operator. 

\begin{figure}[h!t!]
\centering
\includegraphics[scale=0.4]{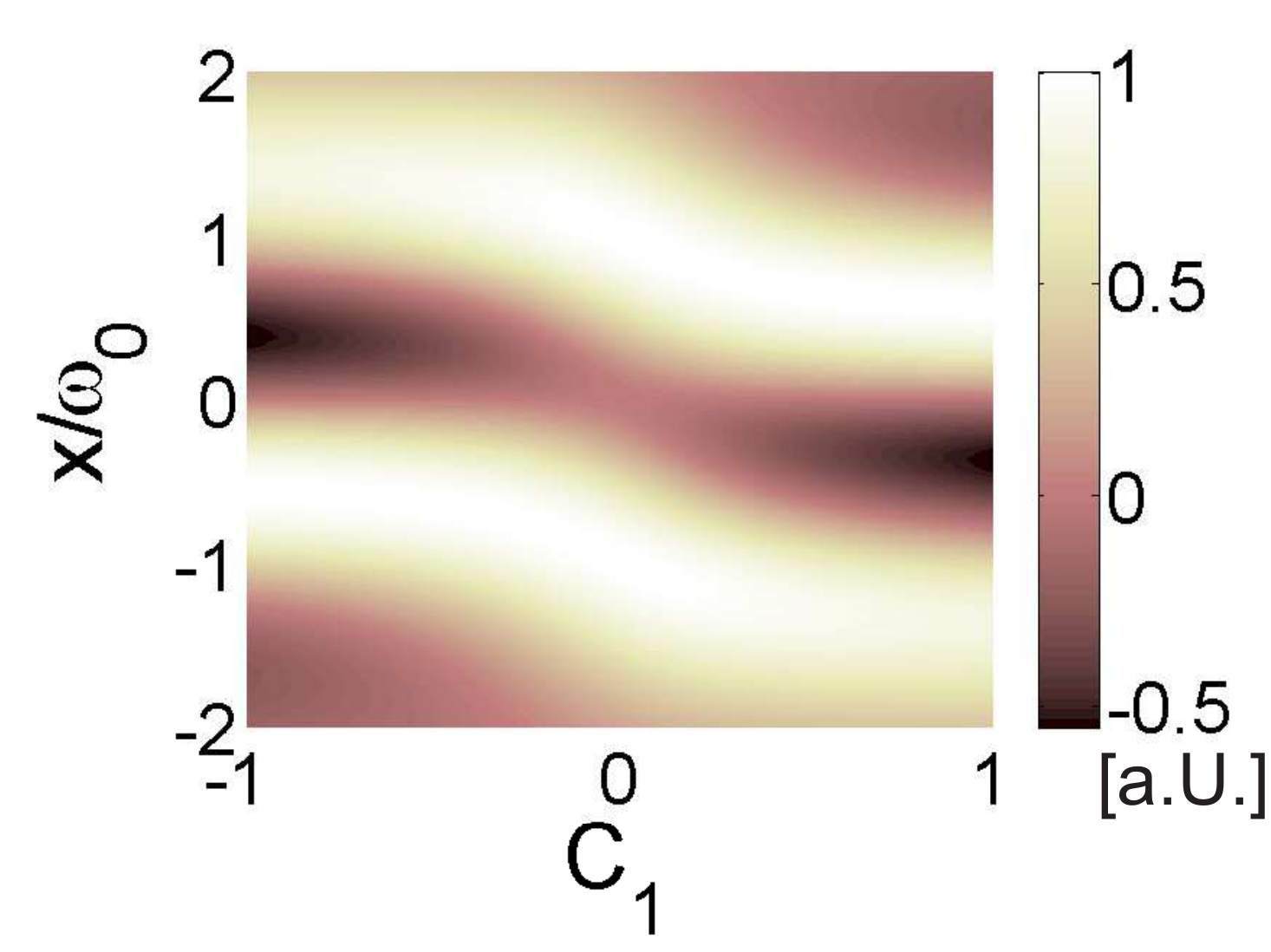}\text{(a)}
\includegraphics[scale=0.4]{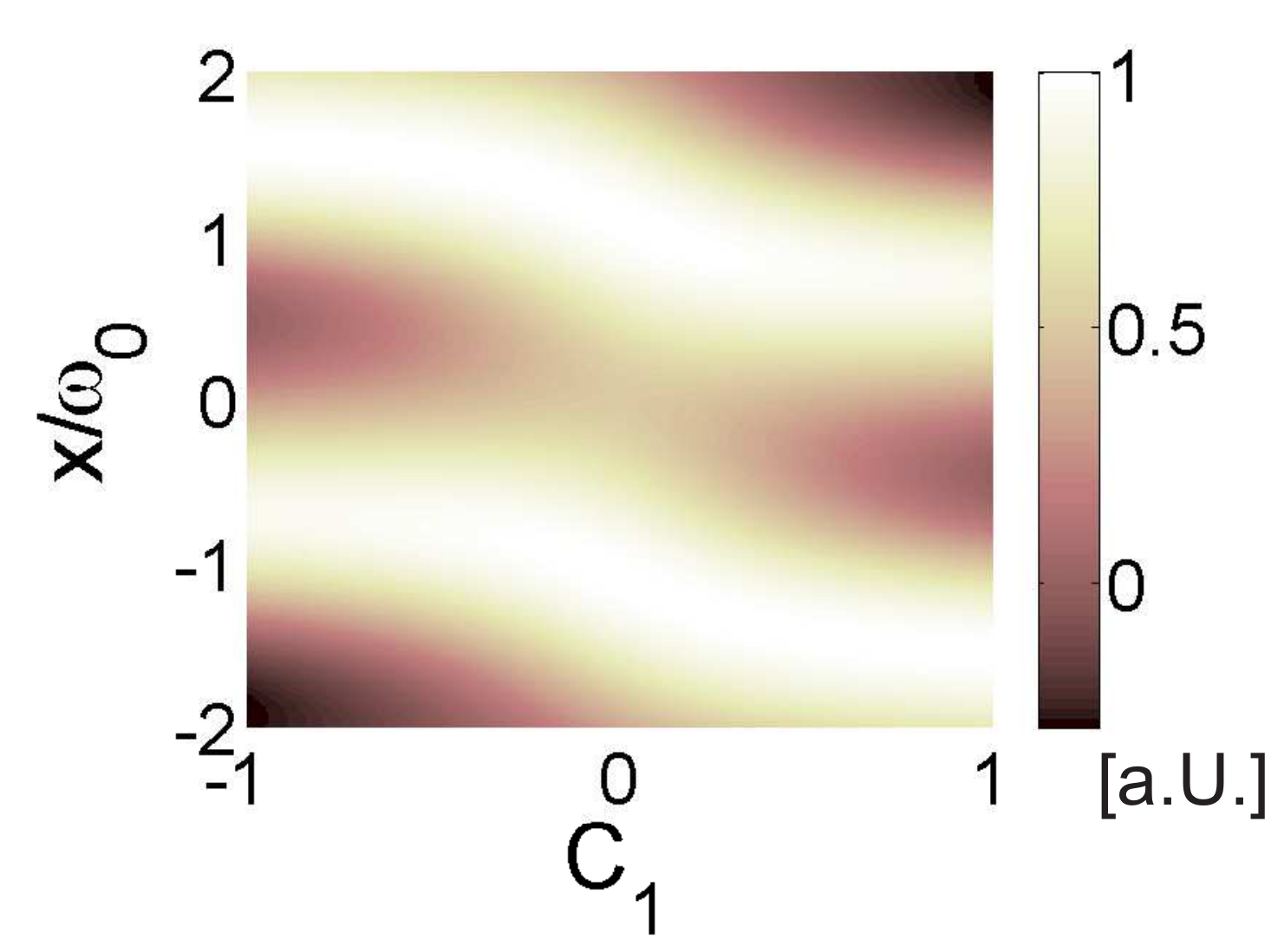}\text{(b)}
\caption{Dependence of calculated beam projections of $x-$ (a) and $y-$ (b) polarized HG$_{10}$ modes on the coefficients $C_1$ according to Eq. (\ref{eq:adopt_dersp}) as influenced by the local electric field.}
\label{fig:exp_setup3}
\end{figure}

The physical meaning behind Eq. (\ref{eq:adopt_der}) is the following. The first term in the sum ($n=1$) is due to the local response of the knife-edge to the $s$- or $p$-polarized electric field and it is mainly associated with the translation operator $U_E(x+\mathrm{d} x)\approx U_E(x)+ \mathrm{d} x \partial U_E(x)/\partial x$. Indeed, if we take either the projection of the $s$-polarized constituents of a radially polarized beam (see Eq. (\ref{eq:sumU})) or a $p$-polarized constituents of an azimuthally polarized beam and plot the resulting beam profiles for various values of $C_1$, we notice a resulting profile which is displaced into the knife-edge or away from it, see Figs. \ref{fig:exp_setup3}(a) and  \ref{fig:exp_setup3}(b). As the expansion coefficient $C_1$ increases, artifacts such as negative values and distortions appear in the resulting profile. We note that coefficients $C_n$ determine the knife-edge as a system and they have to be obtained either experimentally \cite{CHub13} or numerically from the analytical model \cite{PMar11, CHub16}.

We substitute now the expressions of the electric field densities from Eq. (\ref{eq:sumU}) into Eq. (\ref{eq:adopt_der}) and obtain for $s$- and $p$-projections
\begin{align}
	\frac{\partial P}{P_0\partial x_0}=\sum _{m_1,m_2 }B_{m_1,m_2}\left[G_{m_1,m_2}^{(s,p)} + \sum_{n=1}^{\infty} \sum _{l=0}^n\frac{C_n n!}{l!(n-l)!} G_{m_1+1,m_2+n-l}^{(s,p)}\right],
	\label{eq:adopt_dersp}
\end{align}
where
\begin{align}
	G_{m_1,m_2}^{(p)}=g^{(m_1)}_{2}\left( x\right)g^{(m_2)}_{2}\left(x \right), \quad G_{m_1,m_2}^{(s)}=g^{(m_1)}_{1}\left( x\right)g^{(m_2)}_{1}\left(x \right)+\alpha _{m_1,m_2} g^{(m_1+1)}_{1}\left( x\right)g^{(m_2+1)}_{1}\left(x \right).
	\label{eq:GsGp}
\end{align}

It was demonstrated, that for functions with Gaussian envelopes, derivatives up to fourth order are sufficient \cite{CHub13}, thus the inner sum in Eq. (\ref{eq:adopt_dersp}) contains up to $14$ different combinations of HG polynomials for a single unknown  $B_{m_1,m_2}$. One can restrict oneself also in the accuracy of the beam description by lowering upper limits of the indices $m_1$ and $m_2$, thus resulting in a robust fitting algorithm, which can be used for beam reconstruction, provided $C_n$ coefficients were determined numerically or experimentally.

\section{Experimental results and adapted fit algorithm}

\begin{figure}[t!]
\includegraphics[scale=0.75]{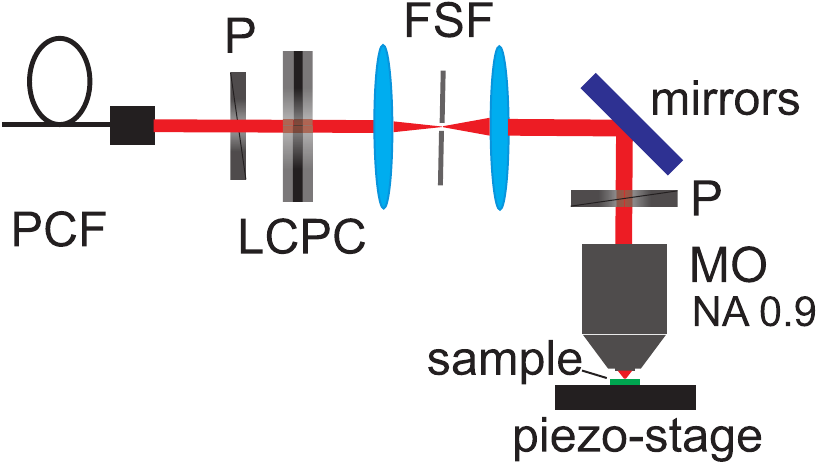}  \text{(a)} 
\includegraphics[scale=0.75]{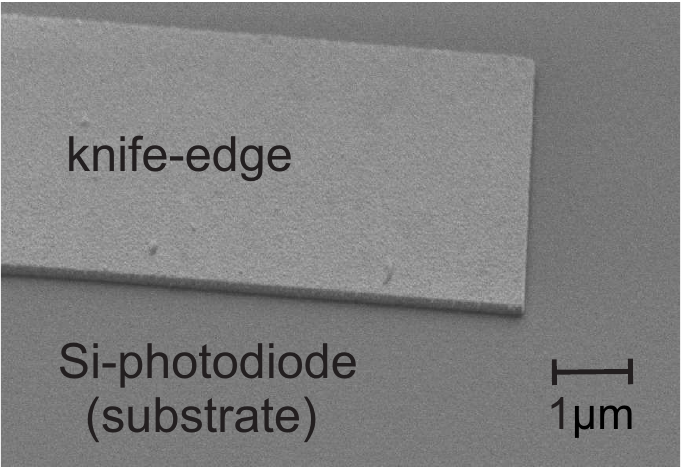} \text{(b)}
\caption {Schematic illustration of the experimental setup (a) (P - Polarizer, LCPC -  liquid crystal polarization converter, FSF - Fourier spatial filter, MO - microscope objective), SEM image of the used knife-edge sample (b) (gold with a thickness $h$ = $70$ nm and a width $d_0$ $\approx$ $3$ $\mu$m fabricated on a silicon photodiode).} 
\label{fig:Bild1}
\end{figure}

In the following section we will briefly discuss the experimental setup, the principle of the measurement and the used knife-edge sample. A detailed discussion of the experimental basics can be found in Refs. \cite{PMar11, CHub16}. The experiments were performed at wavelengths between $535$ nm and $700$ nm using a tunable laser system from TOPTICA. The emitted laser beam is coupled into a photonic crystal fiber (PCF) to obtain a Gaussian beam profile, after collimation this linearly polarized Gaussian beam is converted into a radially or azimuthally polarized mode by a liquid crystal polarization converter (LCPC) (see Fig. \ref{fig:Bild1}(a)) \cite{LCRad}. Afterwards the beam is filtered by a Fourier spatial filter (FSF) that consists of two lenses and a pinhole to achieve a high mode quality. The resulting beam is guided into a high NA microscope objective by a set of mirrors to get focused on the knife-edge sample. For measurements with the fundamental $x$- and $y$-polarized linear constituents of the radially or azimuthally polarized modes, a linear polarizer is used in front of the objective. For the measurements, knife-edges are line-scanned through the focal spot by a piezostage and the power of the light beam that is not blocked by the knife-edge is detected by a photodiode underneath. This way projections of the beam profile are measured as already discussed in section 2.1. 
For these measurements knife-edges made of gold with a thickness $h$ of $70$ nm and a width $d_0$ of $3$ $\mu$m ($\pm$ $50$ nm) are fabricated on silicon (Si) photodiodes as substrate (see Fig. \ref{fig:Bild1}(b)) as they have been already used in \cite{CHub16}. 

\begin{figure}[t!]
\centering
\includegraphics[scale=0.23]{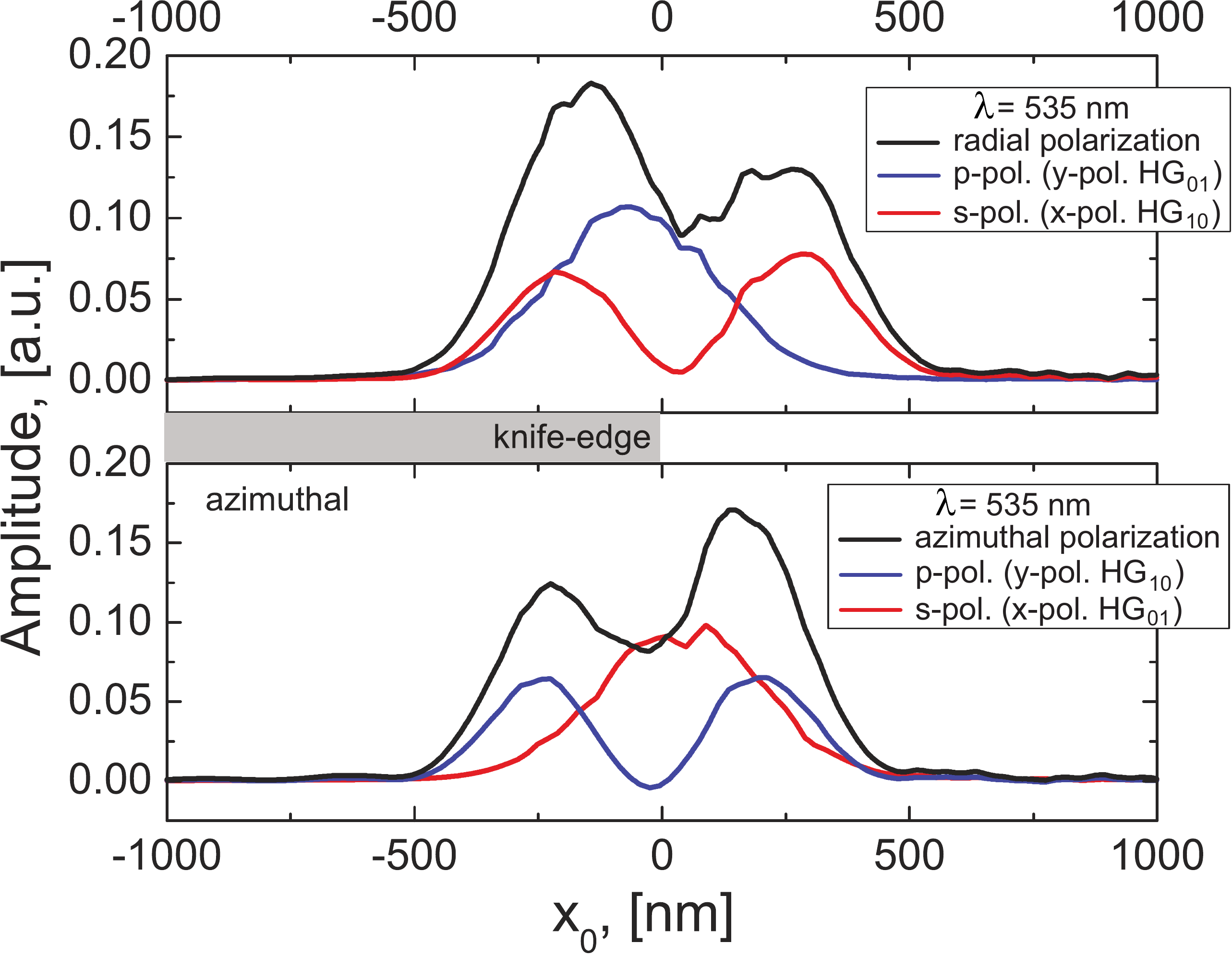} \text{(a)}
\includegraphics[scale=0.23]{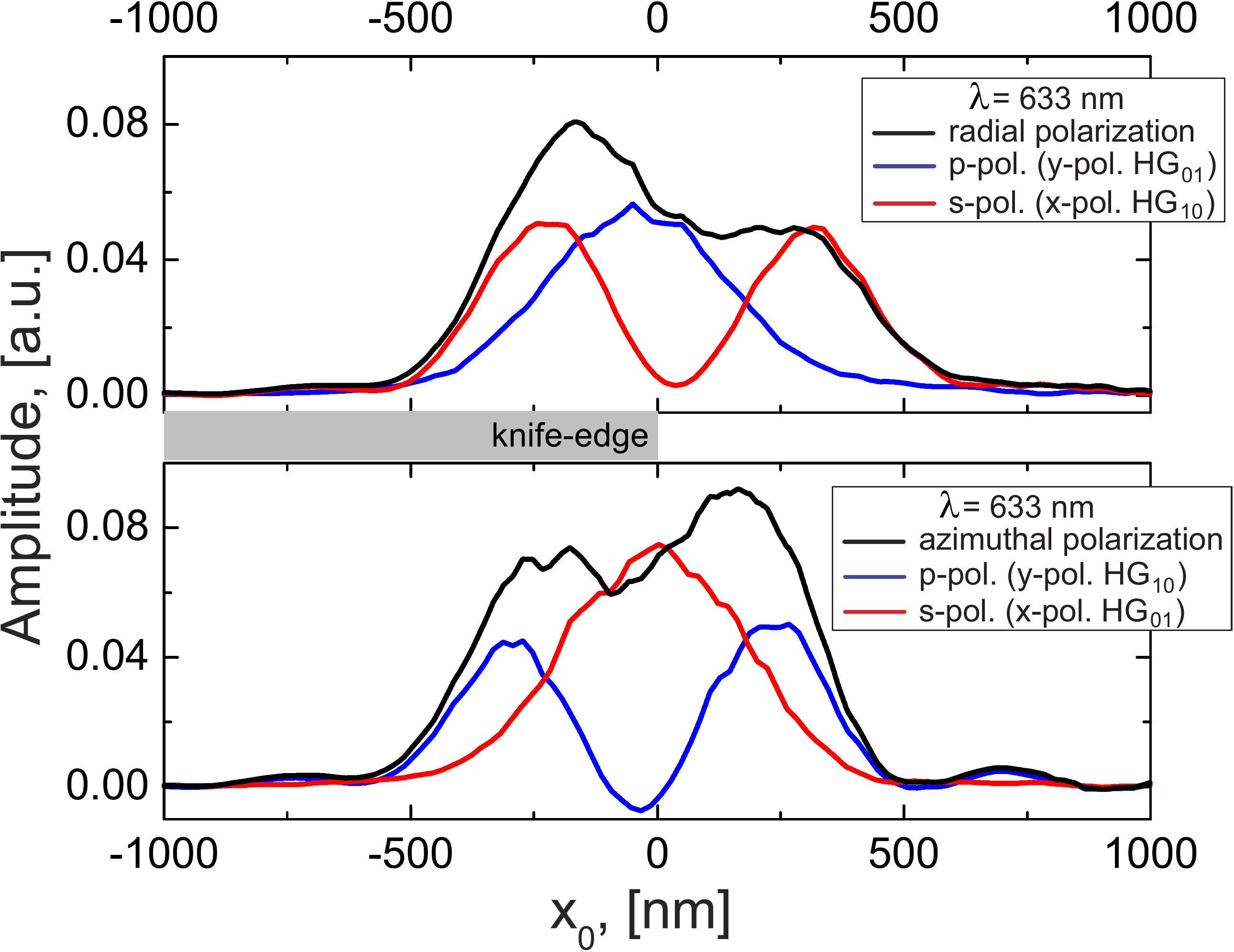} \text{(b)}
\caption{Experimentally measured projections of radially and azimuthally polarized beams (black) and their linear $s$-, $p$- polarized constituents (red, blue) for a) $\lambda=535$ nm, b) $\lambda=633$ nm. The accuracy of determining the actual position of the projections relative to the knife-edge is limited by error in measuring the width $d_0$ of the knife-edge by SEM.}
\label{fig:projections}
\end{figure}

We start the discussion of our experimental results by exemplarily demonstrating the aforementioned artifacts appearing while beam profiling a radially and an azimuthally polarized beam and their linearly polarized constituents at two different wavelengths, see Figs. \ref{fig:projections}(a) and \ref{fig:projections}(b). In all cases, one notices the impact of the wavelength on the shape and relative position of the measured beam projection. While for a particular wavelength of $\lambda=535$ nm the shape of the $p$-polarized linear constituent of an azimuthally polarized beam seems to be sufficiently preserved, see Fig. \ref{fig:projections}(a), further analysis reveals a shift of the profile into the knife-edge and a significantly modified beam width. The projection of the $s$-polarized constituent in this case has a Gaussian shape, its width however is also experiencing changes and its center is shifted away from the knife-edge comparable to the artifacts observed in beam profiling of a linearly polarized Gaussian beams \cite{PMar11}. While the projection of the $p$-polarized constituent of a radially polarized beam is shifted into the knife-edge, the projection of the $s$-polarized constituent of a radially polarized beam is strongly altered in all cases Figs. \ref{fig:projections}(a) and \ref{fig:projections}(b). We find that the contribution of the $z$-component of the electric field in the measured beam profile is in the latter case much weaker than expected as the reconstructed HG$_{10}$ mode shows a minimum close to zero. Therefore the observed profile is more typical for the linearly polarized constituent of a azimuthallly than for a radially polarized beam. The projection of the $p$-polarized constituent of an azimuthally polarized beam experiences not only relative shifts (see Fig. \ref{fig:projections}(b)), but one notices negative values of the projection curve at the center of the beam. This finding is especially surprising, considering the fact that a $p$-polarized HG$_{10}$ mode has no $z$-component in the center of the beam.

\begin{figure}[t!]
\centering
\includegraphics[scale=0.25]{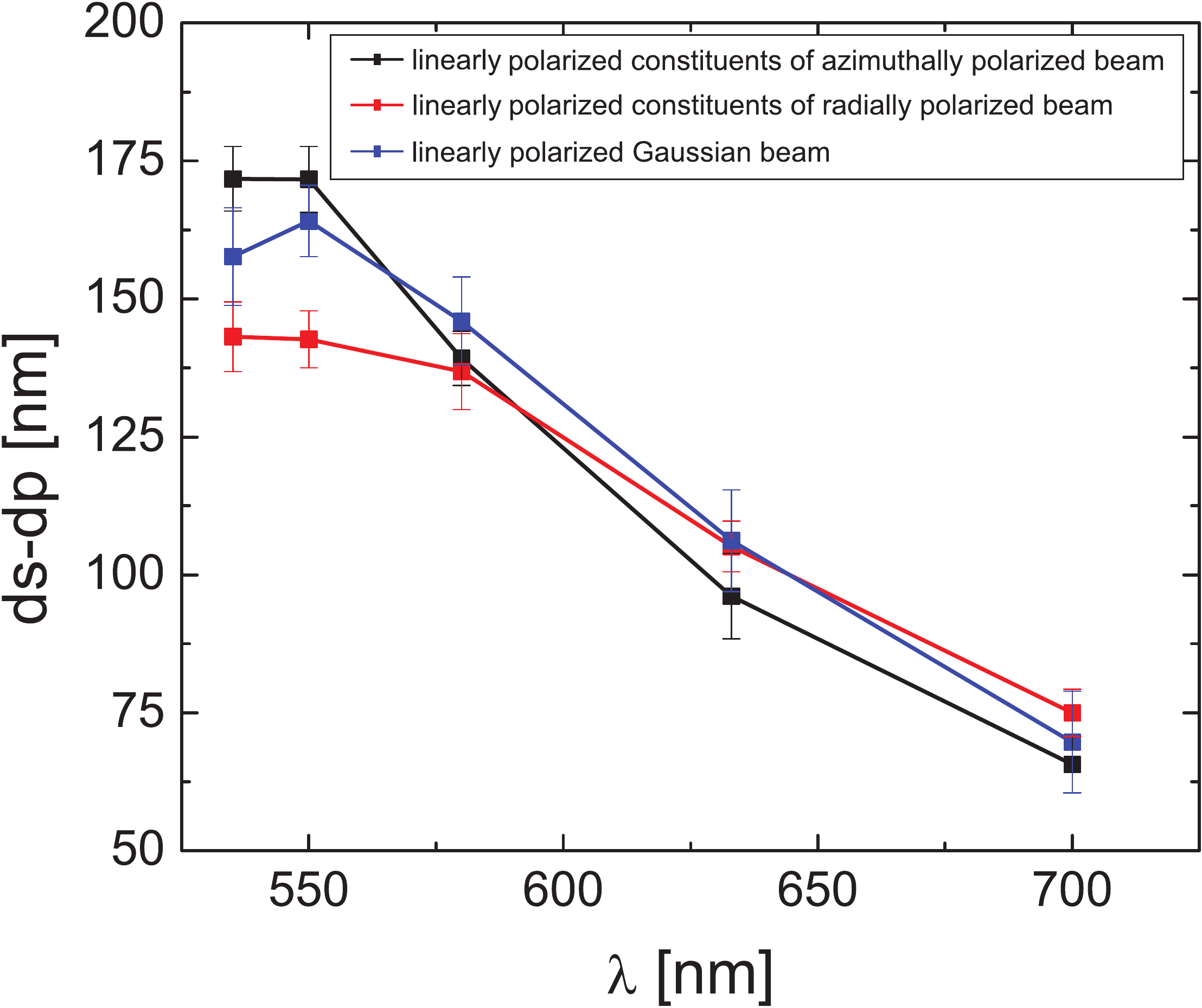}\text{(a)} 
\includegraphics[scale=0.25]{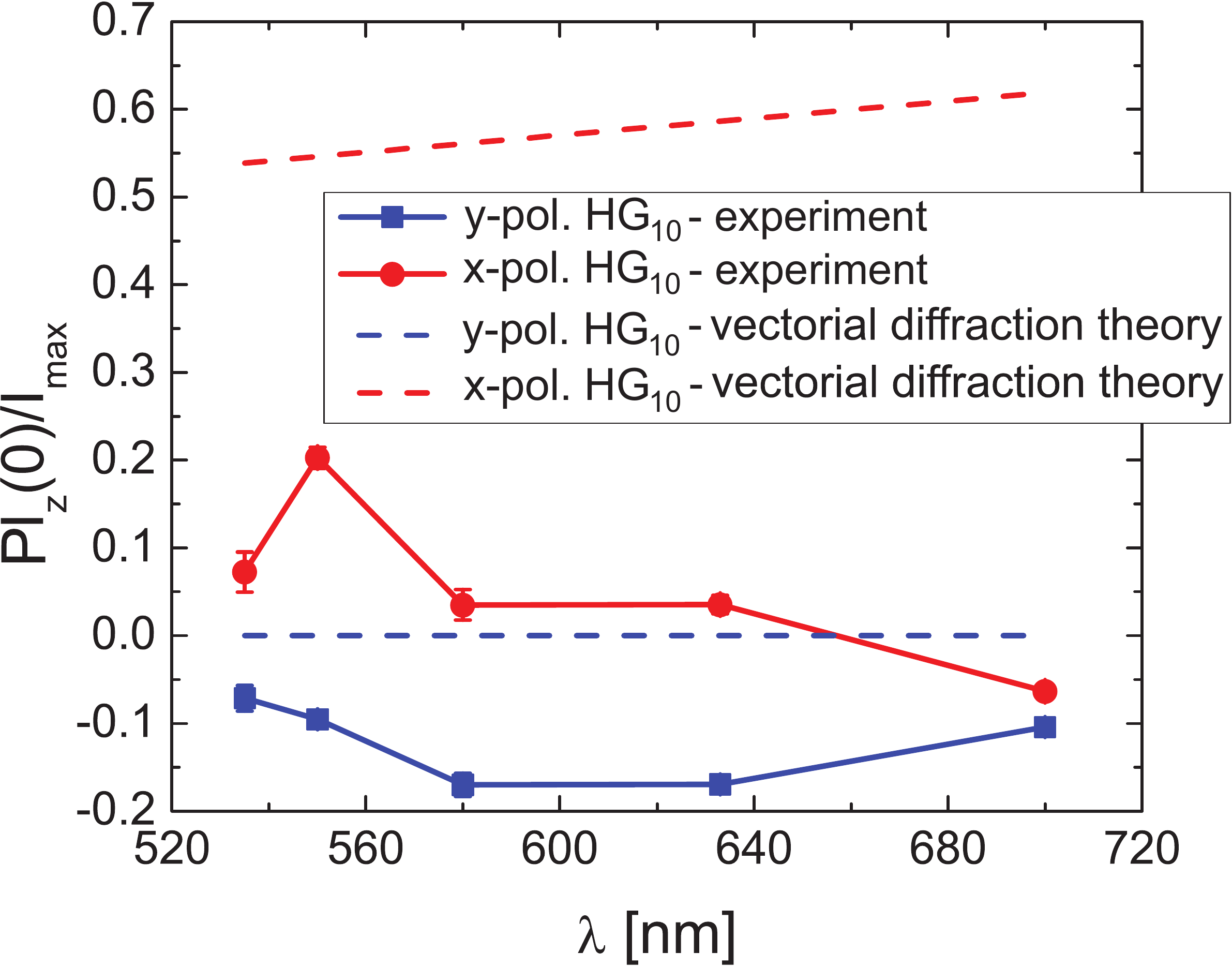}\text{(b)} 
\caption{Experimentally measured relative shift $d_s-d_p$ for linearly, $s$- and $p$-polarized Gaussian beams (black) and for the linearly $s$- and $p$-polarized constituents of radially (red) and azimuthally (blue) polarized beams for different wavelengths $\lambda$ (a). Comparison between the projected intensity (PI) for the $z$-component of the electric field in the center of the measured beam (experiment) and numerically obtained values from vectorial diffraction theory \cite{BRic59} (b).}
\label{fig:radshift}
\end{figure}

As a next step, we have retrieved specific points from the projections of all linearly polarized constituents, indicted in Fig. \ref{fig:exp_setup} and Fig. \ref{fig:exp_setup2} by the red and blue vertical lines, to determine the shift of the projections $d_s$ and $d_p$ from the knife-edges in a systematic way. We found in agreement with theoretical predictions that the projections of $s$- and $p$-polarized constituent of both radially and azimuthally polarized beams experience a similar shifts of their relative positions to the knife-edge meaning that $s$-polarized projections move away from the knife-edge whereas $p$-polarized projections move into the knife-edge. Similar phenomena were recently reported for linearly polarized Gaussian beams, see \cite{CHub16}. Finally, we compare the measured relative shifts $d_s-d_p$ between $s$- and $p$-polarized projections of a linearly polarized Gaussian beam and radially and azimuthally polarized beams, see Fig. \ref{fig:radshift}(a). As expected, their relative shifts are also comparable to each other for different wavelengths, hinting at a experimental verification that coefficients $C_n$ in Eq. (\ref{eq:adopt_der}) depend on the specific knife-edge sample and are independent of the beam shape, which is actually profiled.

At next we compared the measured intensities of the projected linear constituents for the z-component of the electric field at the specific points (see Fig. \ref{fig:exp_setup2}) with those obtained numerically from Debye integrals \cite{BRic59}, see Fig. \ref{fig:radshift}(b). In all cases we have normalized the on-axis intensity at the point $z=0$ to the maximal value of the $s$, $p$- polarized HG$_{10}$ modes. First, we notice, that the conventionally reconstructed linear constituent of the radially polarized beam has on-axis intensities, which are many times smaller than the expected ones. For one particular wavelength the on-axis intensity reaches $20 \%$ of the maximal value, which is almost three times smaller than one might expect from the numerical simulations. At a wavelength of $\lambda= 700$ nm we observe negative values of the projected on-axis intensities. The same is observed for all wavelengths for the reconstructed projections of a HG$_{10}$ mode in the case of azimuthal polarization. In all cases our experimental observations are inline with our expectations from previous sections.

\begin{figure}[ht!]
\centering
\includegraphics[scale=0.22]{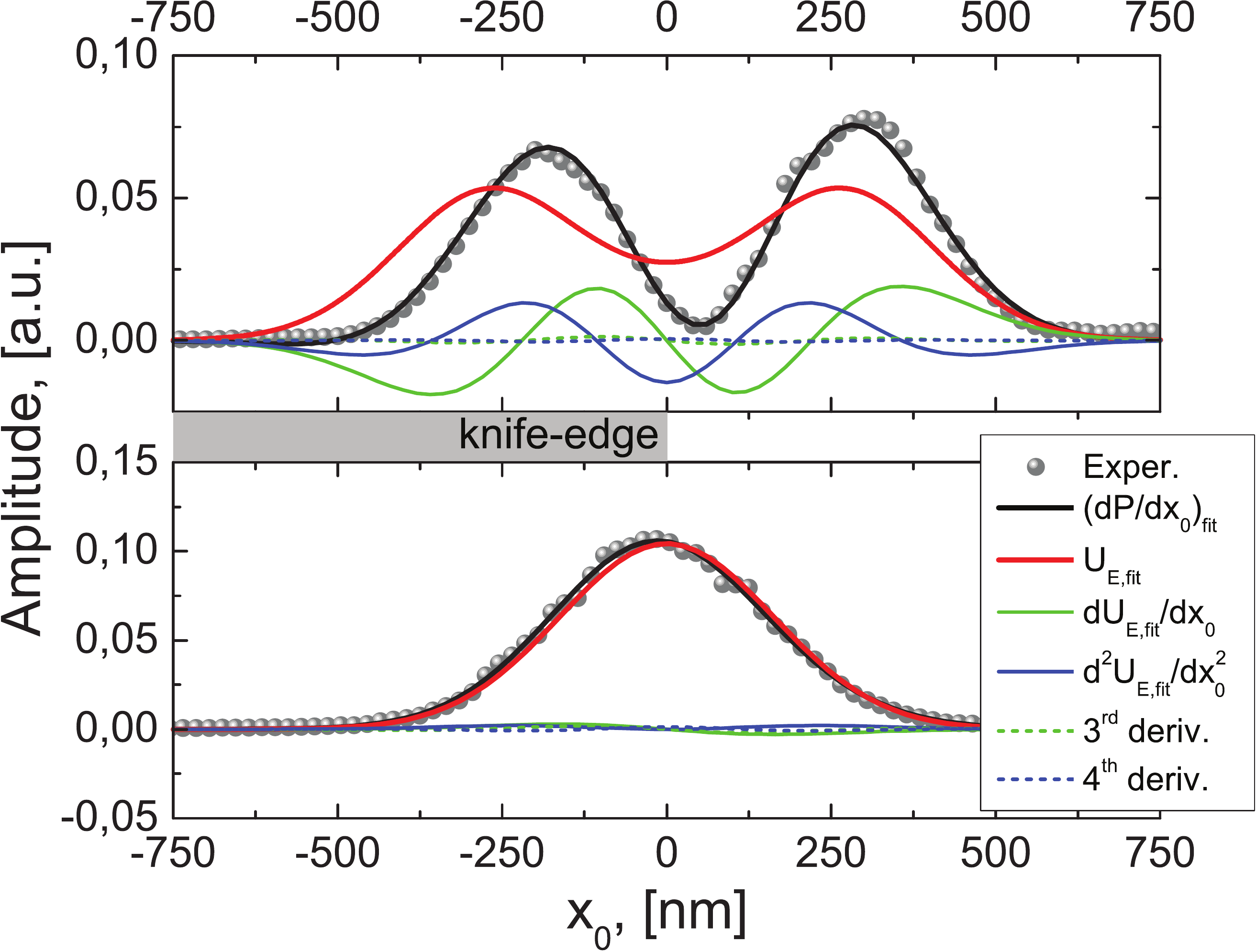}\text{(a)}
\includegraphics[scale=0.22]{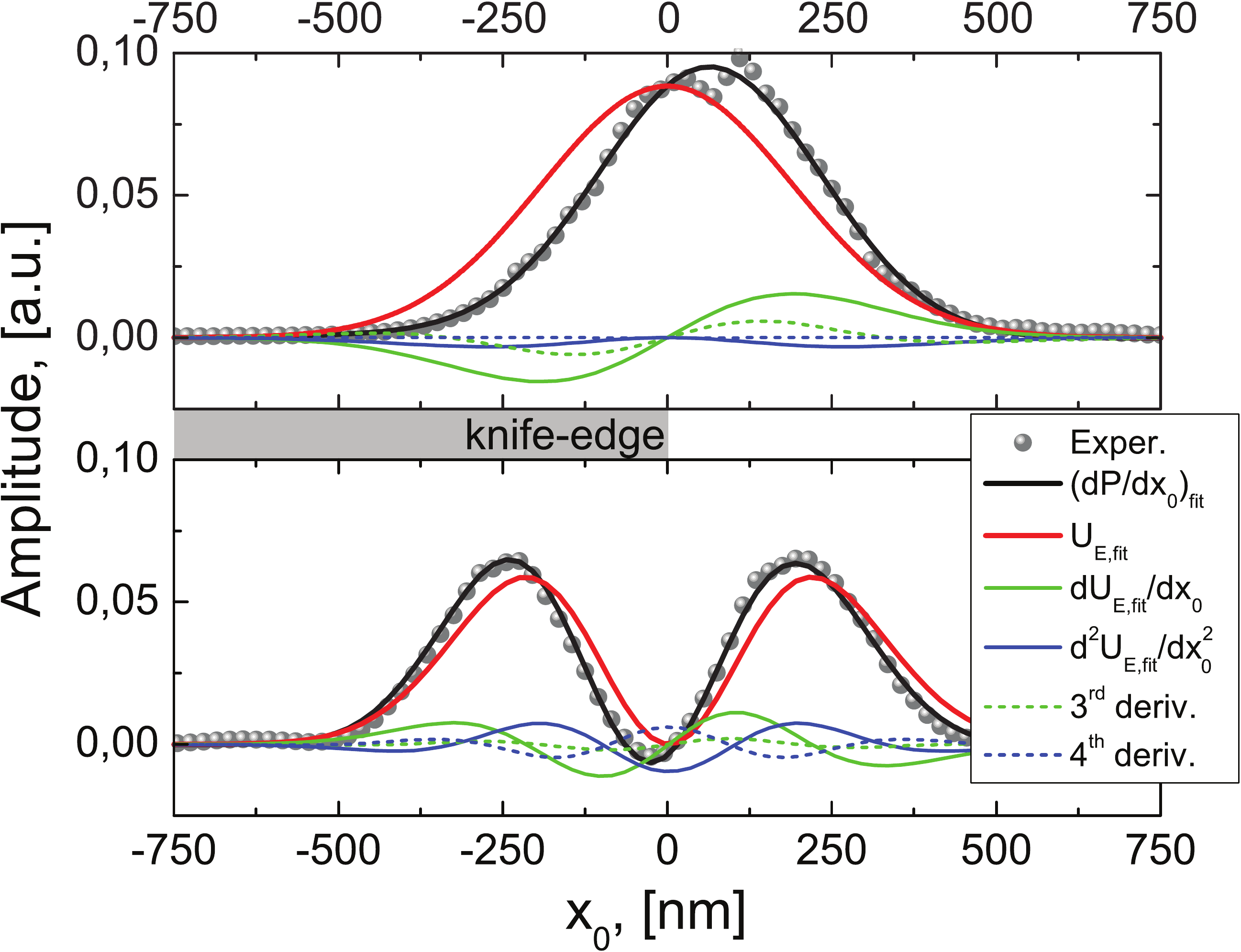}\text{(b)}
\includegraphics[scale=0.22]{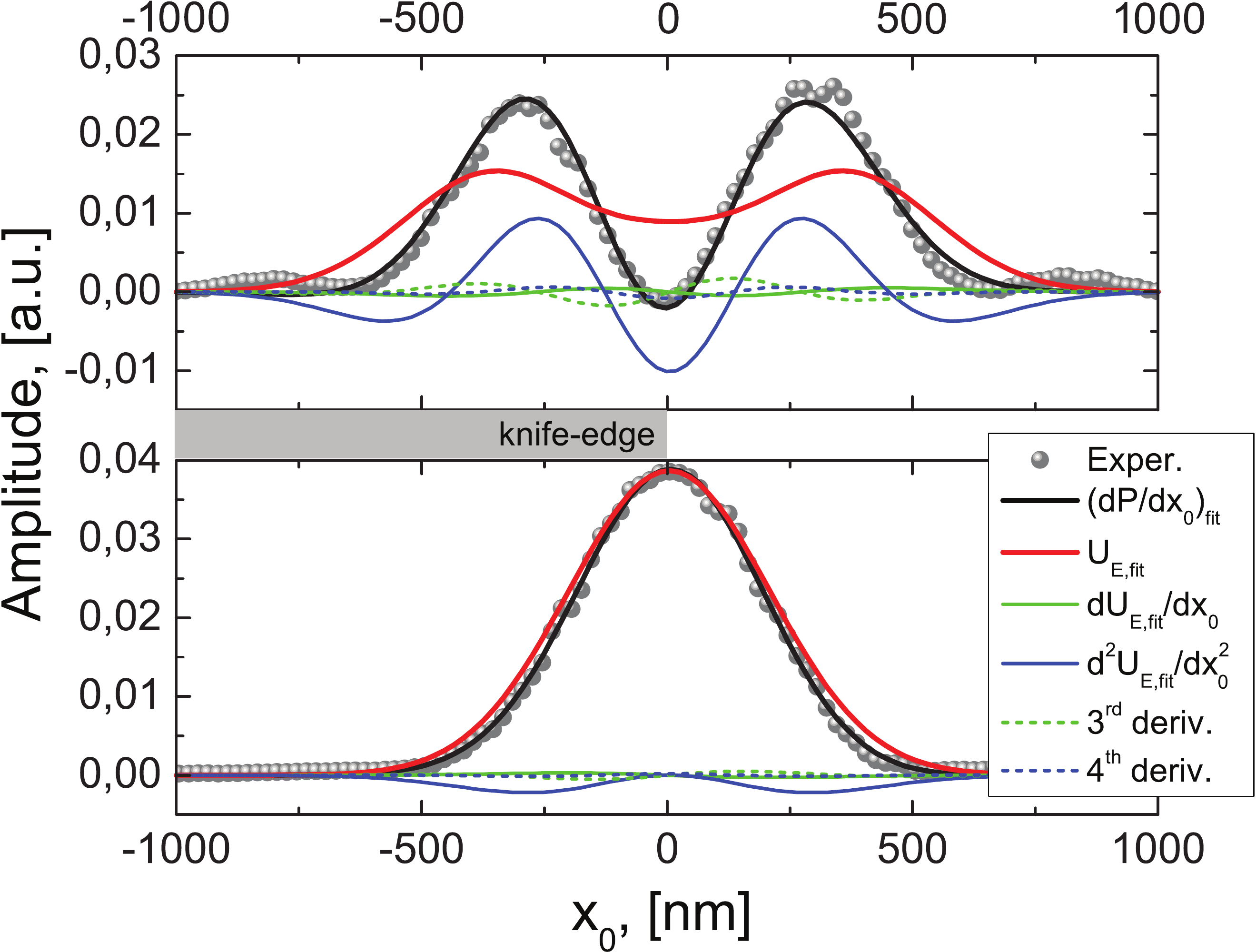}\text{(c)}
\includegraphics[scale=0.22]{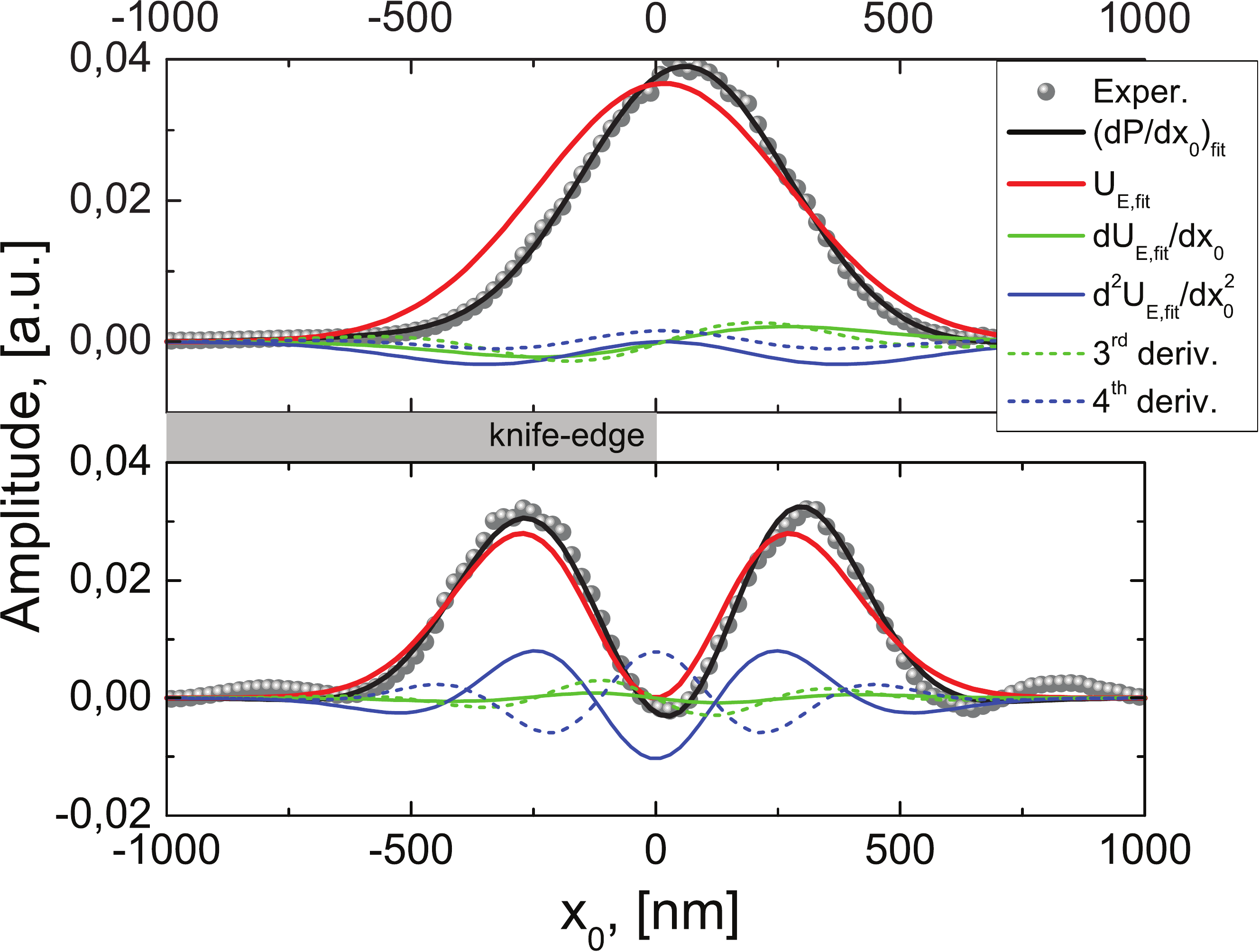}\text{(d)}
\caption{Depiction of the adapted knife-edge method for linear constituents of  tightly focused radially (a, c) and azimuthally (b, d) polarized beams. The derivatives of the experimentally measured photocurrents (gray circles) and the fitted curve (black) with the beam profile (red) and its first four derivatives are shown for $\lambda = 535$ nm (a,b) and $\lambda= 700$ nm (c,d). The states of polarization are $s$ (first row) and $p$ (second row).}
\label{fig:examp}
\end{figure}  

Lastly, based on the ansatz proposed in Eq. (\ref{eq:adopt_dersp}) we have implemented a least-square fitting algorithm as a proof-of-concept test, where we have restricted ourselves to up to the fourth derivative of the electric field energy density projection $U_{E}(x_0)$ and use calculated beam widths from Debye theory as a fixed parameter. For the sake of simplicity, we have used approximations from Eq. (\ref{eq:sumU}) for the components of the $s$- and $p$-polarized beams in the plane of the projection. In addition the real position of the knife-edge was predetermined before fitting to reduce the number of free parameters in the fitting-routine and to fix the coordinate frame. For that purpose, we experimentally measure the distances $d_{0}$ between both edges using a scanning electron microscope (SEM), find the center $x_c$ between the peak values of both projections in one scan and finally set the actual positions of both knife-edges to be at $x_c \pm d_0/2$. An example of such a fitting procedure is presented in Fig. \ref{fig:examp} for $s$- and $p$-polarized constituents of radially and azimuthally polarized beams and a wavelength of $535$ nm and $700$ nm. It turns out that for all investigated wavelengths between $535$ nm and $700$ nm by simultaneously ensuring $d_s=d_p=d_0$ the fitting algorithm has successfully converged towards realistic beam projections, resulting in a good overlap between the theoretical expectations from vectorial diffraction theory and reconstructed beam profiles.

\section{Conclusions}
In conclusion, we have analyzed the performance of the knife-edge method for highly focused radially and azimuthally polarized beams and their linearly polarized constituents. For the correction of the observed modifications in these knife-edge measurements we presented a straight-forward and easy to implement method that is based on the adapted knife-edge reconstruction scheme. This way we are able to retrieve the beam projections of radially and azimuthally polarized laser beams with respect to their linear constituents, for which shifts and deformations of the reconstructed projections as observed in conventional knife-edge measurements can be corrected for.

\section{Acknowledgments}
We thank Stefan Malzer, Isabel G\"a\ss ner, Olga Rusina and Irina Harder for their valuable support in preparing the samples.
\end{document}